\documentclass[a4paper,USenglish,autoref,cleveref,thm-restate]{lipics-v2021}

\newif\ifextendedversion
\extendedversiontrue

\ifextendedversion
\pdfoutput=1 
\fi

\usepackage{listings}
\usepackage{todonotes}
\usepackage{here,caption}
\usepackage{subfiles}
\usepackage{multicol}
\usepackage{mytikzsettings}

\usepackage{color,xcolor}
\usepackage{etoolbox}
\usepackage{textcomp}

\definecolor{stComment}{rgb}{0.5,0.5,0.5}
\definecolor{stString}{rgb}{0.58,0,0.82}
\definecolor{stKeywords}{rgb}{0.21,0.55,0.7}
\definecolor{stNumbers}{rgb}{.5,0,0}

\newtoggle{InString}{}
\togglefalse{InString}
\newcommand*{\ColorIfNotInString}[1]{\iftoggle{InString}{#1}{\color{stNumbers}#1}}%

\lstdefinelanguage{Smalltalk}{
  keywordstyle=\color{stKeywords},
  commentstyle=\color{stComment},
  stringstyle=\color{stString},
  alsoletter=\#,
  identifierstyle=\ttfamily,
  showstringspaces=false,
  morekeywords={true,false,self,super,nil},
  sensitive=true,
  morecomment=[s]{"}{"},
  morestring=[d]',
  style=SmalltalkStyle,
  tabsize=2,
  upquote=true,
}

\lstdefinestyle{SmalltalkStyle}{
  literate=%
  {0}{{{\ColorIfNotInString{0}}}}1%
  {1}{{{\ColorIfNotInString{1}}}}1%
  {2}{{{\ColorIfNotInString{2}}}}1%
  {3}{{{\ColorIfNotInString{3}}}}1%
  {4}{{{\ColorIfNotInString{4}}}}1%
  {5}{{{\ColorIfNotInString{5}}}}1%
  {6}{{{\ColorIfNotInString{6}}}}1%
  {7}{{{\ColorIfNotInString{7}}}}1%
  {8}{{{\ColorIfNotInString{8}}}}1%
  {9}{{{\ColorIfNotInString{9}}}}1%
}

\lstdefinelanguage{myPython}{
    keywords={def, return, if, elif, else, for, while, break, continue, pass,
      import, from, as, class, try, except, raise, with, lambda, yield, assert,
      async, await, nonlocal, global, del, in, is, and, or, not,
    },
    keywordstyle=\color{blue}\bfseries,
    keywordstyle=[2]{\color{red}\bfseries},
    ndkeywords={True, False, None},
    ndkeywordstyle=\color{magenta}\bfseries,
    identifierstyle=\color{black},
    sensitive=true,
    comment=[l]{\#},
    commentstyle=\color{gray}\ttfamily,
    stringstyle=\color{stString},
    morestring=[b]',
    morestring=[b]",
    morecomment=[s]{""" }{ """},
    morecomment=[s]{''' }{ '''},
    morekeywords=[2]{@enable\_threaded\_code, @enable\_shallow\_tracing,
      @dont\_look\_inside, dummy}
    escapechar={!},
}

\newcommand{\jit}{JIT}
\newcommand{\vm}{VM}
\newcommand{\vms}{VMs}
\newcommand{\rpython}{RPython}
\newcommand{\truffle}{Graal/Truffle}
\newcommand{\tcg}{meta-compiler--based threaded code generation}

\title{A Lightweight Method for Generating Multi-Tier JIT Compilation Virtual Machine in a Meta-Tracing Compiler Framework}
\titlerunning{A Lightweight Method for Generating Multi-Tier JIT Compilation VM in a ..} 

\author{Yusuke Izawa}
{Tokyo Metropolitan University, Japan}
{yizawa@acm.org}
{https://orcid.org/0009-0005-4170-9977}
{}

\author{Hidehiko Masuhara}
{Institute of Science Tokyo, Japan}
{masuhara@acm.org}
{https://orcid.org/0000-0002-8837-5303}
{}

\author{Carl Friedrich Bolz-Tereick}
{Heinrich-Heine-Universit\"at D\"usseldorf, Germany}
{cfbolz@gmx.de}
{https://orcid.org/0000-0003-4562-1356}
{}

\authorrunning{Y.Izawa et al.} 

\Copyright{Yusuke Izawa, Hidehiko Masuhara, and Carl Friedrich Bolz-Tereick}



\category{} 

\ifextendedversion
\else
\relatedversion{} 
\relatedversiondetails[cite={ecoop2025extended}]{Full Version}{http://arxiv.org/abs/2504.17460}
\fi

\supplement{}

\supplementdetails[subcategory={Source Code}, cite={}, swhid={}]{Software}{https://doi.org/10.5281/zenodo.15286979}
\supplementdetails[subcategory={Source Code}, cite={}, swhid={}]{Software}{https://doi.org/10.5281/zenodo.15286954}
\supplementdetails[subcategory={Source Code}, cite={}, swhid={}]{Software}{https://doi.org/10.5281/zenodo.15287001}

\funding{%
This research was supported by KAKENHI grant nos.\ 25K21180, 24K23866, 23H03368,
and 23K28058, and JST ACT-X grant no.\ JPMJAX2003.}

\acknowledgements{%
  We would like to thank Stefan Marr for his valuable advice on the measurement
  methodology and the evaluation results of 2SOM. We are also grateful to the
  anonymous reviewers for their insightful comments, which helped improve this
  manuscript.}  

\nolinenumbers 

\EventEditors{Jonathan Aldrich and Alexandra Silva}
\EventNoEds{2}
\EventLongTitle{39th European Conference on Object-Oriented Programming (ECOOP 2025)}
\EventShortTitle{ECOOP 2025}
\EventAcronym{ECOOP}
\EventYear{2025}
\EventDate{June 30--July 2, 2025}
\EventLocation{Bergen, Norway}
\EventLogo{}
\SeriesVolume{333}
\ArticleNo{16}

\begin{document}

\maketitle

\begin{abstract}
Meta-compiler frameworks, such as RPython and Graal/Truffle, generate high-performance virtual machines (VMs)
from interpreter definitions. Although they generate VMs with high-quality just-in-time (JIT) compilers,
they still lack an important feature that dedicated VMs (i.e., VMs that are developed for specific languages)
have, namely \emph{multi-tier compilation}.  Multi-tier compilation uses light-weight compilers at early stages
and highly optimizing compilers at later stages in order to balance between compilation overheads and
code quality.

We propose a novel approach to enabling multi-tier compilation in the VMs generated by a meta-compiler framework.
Instead of extending the JIT compiler backend of the framework, our approach drives an existing
(heavyweight) compiler backend in the framework to quickly generate unoptimized native code by merely embedding
directives and compile-time operations into interpreter definitions.

As a validation of the approach, we developed 2SOM, a Simple Object Machine with a two-tier JIT compiler
based on RPython.  2SOM first applies the tier-1 threaded code generator
that is generated by our proposed technique, then, to the loops that exceed a threshold, applies the
tier-2 tracing JIT compiler that is generated by the original RPython framework.
Our performance evaluation that runs a program with a realistic workload
showed that 2SOM improved, when compared against an RPython-based VM, warm-up performance by 15\%,
with merely a 5\% reduction in peak performance.
\end{abstract}

\section{Introduction}
\label{sec:intro}

A \emph{meta-compiler framework}~\cite{bolz2015408,wurthinger:2017:ppe:3062341.3062381} is a
system that generates high-performance virtual machines (\vm{}s) from interpreter definitions.
Traditionally, \vm{}s were manually implemented by \vm{} developers from scratch, including
interpreters, just-in-time (\jit{}) compilers, and garbage collectors. However, by using a
meta-compiler framework, it is possible to generate a \vm{} of comparable quality to traditional
\vm{}s simply by defining the interpreter.  Currently, there are two frameworks in this domain:
RPython~\cite{bolz:2009} and \truffle{}~\cite{wurthinger:2012}. These frameworks have
demonstrated their practical effectiveness by generating high-performance VMs such as
PyPy~\cite{rigo:2006:10.1145/1176617.1176753}, Pycket~\cite{pape:2015:10.1145/2816707.2816716},
TruffleSqueak~\cite{niephaus:2019:gts:3357390.3361024}, and TruffleRuby~\cite{wurthinger:2017:ppe:3062341.3062381}.


Given the importance of multi-tier \jit{} compilation \vm{}s, efforts have been made
with \truffle{} to enable a meta-compiler framework to perform multi-tier \jit{} compilation~\cite{prokopec:2021:graalmulti}.
The multi-tier \jit{} compilation \vm{} has multiple execution tiers
using different compilation methods or optimization levels to balance code quality
and compilation time~\cite{arnold:2000:10.1145/353171.353175}. In \truffle{}'s
methodology, a single \jit{} backend is used, and first-tier and second-tier compilations are
achieved by gradually enabling or disabling optimizations. However, it has not yet achieved the
incorporation of tiers with fundamentally different compilation methods, as has been done
in traditional \vm{}s.

Furthermore, as with traditional \vm{}s, enabling a meta-compiler framework to generate a
multi-tier \jit{} compilation \vm{} requires a significant development effort. In traditional \vm{}s,
developing a multi-tier \jit{} compilation \vm{} is more complex than simply implementing
multiple compilers. At the least, a \vm{} developer develops multiple compilers
with different compilation behaviors---a compiler for rapid compilation speed but generating code
with moderate quality, and a compiler for generating quality code---plus a profiler to
gather runtime information for switching between different compilers, and a mechanism to
transition between code generated by different compilers. Therefore, a framework developer needs to
implement these components at the level of a meta-compiler framework.


To address the implementation cost,  we propose a novel methodology that enables a meta-compiler
framework to generate a multi-tier \jit{} compilation \vm{} without requiring the development of new \jit{}
compilation backends.
Our methodology treats interpreter definitions not only as semantic specifications of the source
language but also as specifications of compilation strategies of the meta-compiler.
Under our methodology, a \vm{} developer provides interpreter definitions for each compilation tier.
Then, \rpython{} can automatically generate the necessary components for multi-tier \jit{} compilation.

Specifically, the proposed methodology:

\begin{description}
\item [(\textbf{Multiple compilers}:)] generates a lightweight compiler for rapid
  code generation during the warm-up phase of a source program, and a heavyweight compiler for
  optimizing the hot spots of a source program,
\item [(\textbf{Profiling mechanism}:)] generates profiling code fragments that dynamically identify
  hot spots and trigger tier transitions, and
\item [(\textbf{Transition mechanism}:)] switches between code generated by different compilers.
\end{description}

To validate our methodology, we implemented a two-level Simple Object Machine, which we call 2SOM for short,
on top of the \rpython{} framework.  2SOM is an extended version of PySOM~\cite{pysom}
that is an implementation of Simple Object Machine~\cite{haupt:2010:sfv:1822090.1822098}
by \rpython{}. 2SOM has two JIT compilers:
\begin{description}
\item[(Tier-1:) A threaded code generator,] which quickly generates subroutine-threaded code for frequently invoked methods.  The threaded code generator is generated by using the \rpython{}'s mechanism based on the seminal ideas of Izawa et al.~\cite{izawa:2022:threadedcodegen} along with several techniques proposed in this paper to make it practical.
\item[(Tier-2:) A tracing \jit{} compiler,] which is generated by \rpython{}~\cite{bolz:2009,gal2009,gal:2006:10.1145/1134760.1134780} and compiles traces with aggressive optimizations to frequently executed loops in the threaded code.  
\end{description}

Our approach effectively enables multi-tier \jit{} compilation for meta-compiler
frameworks. Although this paper demonstrates an application to SOM, the proposed
approach should also be applicable to other \rpython{}-based language implementations.

To evaluate the effectiveness of the proposed approach, we analyzed the performance of 2SOM
with respect to (1) improvement of the warm-up performance for a realistic workload, (2)
degradation of peak performance, and (3) quality of the
threaded code generator. Our evaluation demonstrates that (1) two-level compilation
runs about 15\% faster than tracing \jit{}-only compilation, (2) the degradation is up to
about 5\% compared to tracing \jit{}-only compilation, and (3) our contribution improves
the code quality of about 10\% compared to interpreter execution.



This paper makes the following contributions:

\begin{itemize}
\item A novel approach for creating a multi-tier \jit{} compilation \vm{} using the RPython
  framework by extending interpreter definitions.
\item The optimization of meta-compiler--based threaded code
  generation~\cite{izawa:2022:threadedcodegen}, addressing practical issues and proposing
  solutions for integration with meta-tracing \jit{} compiler frameworks.
\item The implementation of 2SOM, the first \vm{} with two \jit{} compilers---a threaded code
  generator and a tracing compiler---generated entirely from interpreter definitions.
\item A methodology for synthesizing large-scale application programs to evaluate the
  warm-up performance of the \jit{}-compiler-based \vms{}.
\item The improvement of warm-up execution speed in 2SOM for synthesized application programs.
\end{itemize}

The remainder of this paper is organized as follows. Section~\ref{sec:background}
introduces the techniques underlying our proposal, including \jit{} and multi-tier \jit{} compilation,
meta-compilation, meta-tracing compilation, and meta-compiler--based threaded code generation.
Section~\ref{sec:problem}
discusses the challenges of realizing a multi-tier \jit{} compilation \vm{} within a meta-compiler
framework. Section~\ref{sec:proposal} presents the architecture and technical details of
2SOM. Section~\ref{sec:opt_threaded_code} identifies practical problems in meta-compiler-based
threaded code generation and proposes solutions. Section~\ref{sec:evaluation} evaluates
2SOM's performance and discusses the results. Section~\ref{sec:relatedwork}
sets the results into the context of related work, and Section~\ref{sec:conclusion} concludes the paper.

\section{Background}
\label{sec:background}

In this section, we demonstrate the overview and implementation challenges in a multi-tier \jit{}
compiler. Subsequently, we present an overview of the meta-compiler framework. Finally, we
introduce meta-compiler--based threaded code generation, a promising technique for
lightweight compilation. Finally, we provide an overview of the Simple Object Machine that we
used to implement our proposal.

\subsection{JIT and Multi-Tier JIT Compilation}

In general, most \vm{}s have interpreters as their first execution tier.
This is because the interpreter is easy to implement and can perform profiling tasks
such as type information collection. For many languages, a \vm{} developer can likely
develop a lightweight \jit{} compiler with moderate effort. A lightweight \jit{} compiler
generates code quickly, but the performance of the generated code is moderate.
To achieve the desired performance, heavyweight \jit{} compilers are
implemented. The code that they generate is highly optimized but compilation
consumes markedly more time than lightweight \jit{} compilation.

Most traditional \vm{}s such as Java and JavaScript
\vm{}s~\cite{paleczny2001java,kotzmann:2008:10.1145/1369396.1370017,openj9,googlev8,jsc}
use multi-tier \jit{} compilation~\cite{arnold:2000:10.1145/353171.353175}.
A multi-tier \jit{} compilation \vm{} involves a hierarchy of \jit{} compilers, each
offering different optimization tiers. As the optimization tiers grow in number, compilation
time also increases, and the resulting code achieves higher performance.
Most multi-tier \jit{} compilation \vm{}s include at least two lightweight and heavyweight
compilers.

Multi-tier \jit{} compilation dynamically profiles the runtime information of a source program to
select the most appropriate compiler for compilation and execution. Additionally, it
transitions execution to higher optimization levels when beneficial, thereby enhancing the
overall \jit{} performance.

The early execution phase, evaluated by \emph{warm-up
performance}, and the computationally intensive phase, evaluated by \emph{peak performance}, are
crucial for improving the \vm{}'s overall performance. The lightweight compiler is
typically used to accelerate warm-up, whereas the heavyweight compiler focuses on maximizing
peak performance. Below, we discuss the relationship between \vm{} execution, warm-up
performance, peak performance, and multi-tier \jit{} compilation in detail.

\subsubsection{Warm-up and Peak Performance in Multi-Tier JIT Compilation}

Warm-up performance refers to the efficiency during the phase required for a \vm{} to
transition from its initial execution. This phase typically involves interpretation or
lightweight compilation to a state where
the optimized machine code generated by the \jit{} compiler is actively used. This transition
is critical for interactive or short-lived applications, for which execution speed during the
early stages of program execution exerts a significant impact on user experience.
For long-running server applications, however, warm-up performance is less critical than
peak performance.

Peak performance, conversely, represents the maximum level of processing power and
resource utilization that a \vm{} can achieve under optimal conditions. This phase is usually
reached after the warm-up phase, when the \vm{} is executing fully optimized code. This is particularly
important for long-running applications, such as server processes, for which sustained high
performance is essential. The C2 compiler in HotSpot JVM~\cite{paleczny2001java} is designed to maximize
peak performance by leveraging the profiling data gathered during the warm-up phase to
generate highly optimized machine code. However, achieving peak performance comes with
trade-offs including the overhead of profiling, the complexity of code transitions, and
potential inefficiencies caused by suboptimal profiling parameters or compilation
strategies.

In the case of HotSpot JVM, the C1 compiler~\cite{kotzmann:2008:10.1145/1369396.1370017}
generates moderately optimized code quickly,
allowing the application to execute at a reasonable speed while gathering profiling
information. This profiling data is then used by the C2 compiler~\cite{paleczny2001java}
to produce highly optimized code, improving performance over time. However, the extra overhead
introduced by tiered compilation, such when as profiling and transitioning between
different levels of compiled code, can undermine warm-up performance if not
carefully managed.

Multi-tier \jit{} compilation technique ensures that frequently executed code is greatly
optimized and achieves faster peak performance. On the other hand, the code parts that are not
frequently executed benefit from quicker compilation,
leading to the improvement of warm-up performance. Correspondingly, overall
execution performance is improved by tailoring the level of optimization to the usage
frequency of each part of the code.

\subsection{Meta-Compilation}

While interpreters are easy to implement, understand, and extend, implementing a
\jit{} compiler is an error-prone task and requires engineering work-hours summing, at times, to dozens of years.

Meta-compilation reduces the amount of work in implementing an individual \jit{} compiler
by making the \jit{} compiler reusable and language-independent. A meta-compiler
compiles the source program together with the interpreter that runs it. By using both the
source code and the interpreter for compilation, meta-compilers can produce highly
optimized code that is equivalent to state-of-the-art \jit{} compilers.

In general, a meta-compiler compiles a source program together with an interpreter that
executes the source program. The process involves the following steps:

\begin{enumerate}
\item \textbf{Hot Spot Identification}: identifying hot spots by executing the
  interpreter,
\item \textbf{Behavior Extraction}: extracting the selected hot spot's behavior, and
\item \textbf{Native Code Translation}: translating this
behavior into native code.
\end{enumerate}

The current meta-compilation system differs from the method of extracting the hot spot's
behavior.  The meta-tracing \jit{} compilation system in \rpython{}~\cite{bolz:2009} traces the
interpreter as it executes the hot spot, whereas the self-optimizing interpreter in
\truffle{}~\cite{wurthinger:2012} applies partial evaluation to it.

Next, we introduce \rpython{}, which we use as the basis of our proposal.

\subsubsection{RPython: A Meta-Tracing Compiler Framework}

While traditional \jit{} compilers compile a frequently executed method as their
compilation unit, tracing \jit{} compilers~\cite{bala:2000,gal2009,bebenita:2010} compile
the execution path of a source program called \emph{trace}. Typically, tracing \jit{}
compilers compile a loop in a source program.

The RPython framework~\cite{rigo:2006:10.1145/1176617.1176753,bolz:2009} uses tracing
\jit{} compilation for its meta-compilation technique. A meta-tracing \jit{} compiler doesn't trace a source program directly but instead traces the executing interpreter.
The \rpython{} framework provides annotations to indicate which part of the interpreter
can be a compilation target. For instance, to find a loop in a source program, an
annotation is inserted where the back edge of the loop occurs in the interpreter. The
meta-tracing compiler starts to trace there when a profiling counter on a back edge exceeds a
threshold. Then, tracing continues until it reaches the inserted annotation
again. The resulting trace is compiled into machine code, and the subsequent executions
of the compiled source program are run in the generated machine code.

To further improve performance, \vm{} developers can use annotations to convey
more information to the meta-tracing \jit{} compiler. For instance, \vm{} developers can specify
that certain variables in the interpreter implementation should be constant, allowing for more aggressive
optimizations. In addition, by specifying a function as side-effect--free, the meta-tracing \jit{}
compiler can replace the function call with a resulting constant if all of the passed arguments
are constant. If not, when the meta-tracing \jit{} compiler sees the same operation on
the same arguments again later, it reuses the result from the previous operation's result.
In addition, if inlining a particular function defined in an interpreter would make the
performance worse, the developers can avoid inlining that function by specifying an
annotation to leave a call instruction on it.

\subsection{Meta-Compiler--Based Threaded Code Generation}

\emph{Threaded code}~\cite{bell:1973:10.1145/362248.362270,dewar75} is
a form of machine code generated from a bytecode-based source program and its
interpreter. Among various types of threading (which include direct
threading~\cite{bell:1973:10.1145/362248.362270} and indirect
threading~\cite{curley:1993a,curley:1993b}, and subroutine
threading~\cite{curley:1993a,curley:1993b}) compiles a sequence of bytecode instructions
into corresponding call instructions to their respective handlers. Threaded code
generation is a lightweight compilation technique, simplifying the process and eliminating
the overhead of bytecode fetching and dispatching~\cite{ertl2003the}.

Meta-compiler--based threaded code generation~\cite{izawa:2022:threadedcodegen}
is a compilation technique for lightweight compilation that generates
subroutine-threaded code using a meta-tracing \jit{} compiler. Unlike other threaded code
generators that develop a compiler from scratch, this approach leverages the interpreter
definitions used in the compilation of the meta-tracing \jit{} compiler. Specifically, it
makes use of the annotation mechanisms provided by the RPython framework. For instance,
annotations are added to prevent inlining within all handler functions in the interpreter
definitions, allowing call instructions to bytecode handlers to be aligned
sequentially in the resulting trace. Additionally, while tracing \jit{}
compilers trace only one side of a branch; this technique introduces mechanisms to trace
all possible paths in the source program in one shot to compile the entire method. In this
way, a new lightweight compilation technique is achieved without creating an entirely new
compilation pipeline within the meta-tracing compiler.

The workflow for threaded code generation is illustrated in Figure~\ref{fig:izawa_et_al_tc_flow}.
First, the threaded code generator traces \verb|strange_add| by executing its bytecode,
as shown in Listing~\ref{lst:som_strange_add_byte}. During this process, it records call
instructions to the corresponding handler functions in the trace. When it encounters a
branch instruction, such as \verb|JUMP_IF L1|, it first traces the false branch. Upon
reaching \verb|RET|, a dummy instruction, \verb|cut_here(return)|, is temporarily
added to the trace. The technique then returns to trace the true branch, stopping at
\verb|RET| again. The resulting trace, shown in Listing~\ref{lst:tc_trace}, includes only
calls to handlers, along with guards, labels, and pseudo instructions such as
\verb|cut_here|.

Next, the technique splits the trace into two parts, \verb|Trace A| and \verb|Trace B|,
replacing the dummy instructions with actual \rpython{} intermediate representations
(e.g., \verb|finish()|). The split traces are shown in
Listing~\ref{lst:tc_after_split}. Finally, it stitches \verb|Trace B| back into
\verb|Trace A| at the guard point, \verb|guard_false(i2)|. This stitched trace
is then compiled into the final assembly code, as displayed in
Listing~\ref{lst:tc_native}.

\begin{figure}[!tp]
  \centering
  \begin{subfigure}[t]{.50\linewidth}
\begin{lstlisting}[language=Smalltalk,caption={Example SOM program.},
label={lst:som_strange_add_source}]
C = (
  calc: n = (
    | x |
    # for ...
    1 to: n do: [ |i|
      x :=
        "Addition: call strange_add"
        x + strange_add: i m: n.
    ] )

  "Newly added method"
  strange_add: n m: m = (
    [ n % 42 == 0 ]
      ifTrue:  [ ^ (m - 42) ]
      ifFalse: [ ^ (n + m) ] )

  "Entry point of this program"
  run = (
    calc: 10000.
  ) )
\end{lstlisting}
  \end{subfigure}\hfill
  \begin{subfigure}[t]{.48\linewidth}
\begin{lstlisting}[
caption=Pseudo bytecode format corresponding to the program shown in
  Listing~\ref{lst:som_strange_add_source}.,
label=lst:som_strange_add_byte
]
calc:
  DUP     # n
  CONST 1 # i
L0:
  CALL "strange_add"
  # ... Same to Figure 2 ...
  RET
strange_add:
  DUP        # n
  CONST 42
  MOD        # n % 42
  JUMP_IF L1
# false branch
  DUP        # n
  DUP        # m
  ADD        # n + m
  RET
# true branch
L1:
  CONST 42
  DUP        # n
  SUB        # n - 42
  RET
\end{lstlisting}
\end{subfigure}
\caption{Example source program and corresponding bytecode representation in 2SOM.}
\label{fig:som_example}
\end{figure}

\begin{figure}[!tp]
\begin{subfigure}[t]{.32\linewidth}
\begin{lstlisting}[caption={Temporal trace from \texttt{strange\_add} traced by
the threaded code generation.}, label={lst:tc_trace}]
# p0 means a red var.
# in interpreter
label(strange_add)
call(DUP, p0)
call(CONST,p0,  42)
i2 = call(MOD, p0)
guard_false(i2)
# false branch
call(DUP, p0)
call(DUP, p0)
call(ADD)
cut_here(return)
# true branch
call(CONST, p0, 42)
call(DUP, p0)
call(SUB, p0)
finish(p0)
\end{lstlisting}
\end{subfigure}\hfill
  \begin{subfigure}[t]{.32\linewidth}
\begin{lstlisting}[
caption={Traces after re-constructing the control flow.},
label={lst:tc_after_split}]
# Trace A
label(strange_add)
call(DUP, p0)
call(CONST, p0, 42)
i2 = call(MOD)
# go to L1 if failed
guard_false(i2) [L1]
# false branch
call(DUP, p0)
call(DUP, p0)
call(ADD, p0)
finish()

# Trace B
# true branch
label(L1)
call(CONST, p0, 42)
call(DUP, p0)
call(SUB, p0)
finish()
      \end{lstlisting}
  \end{subfigure}\hfill
  \begin{subfigure}[t]{.32\linewidth}
\begin{lstlisting}[
caption={Pseudo-assembly code compiled from
Listing~\ref{lst:tc_after_split}.},
label={lst:tc_native}]
strange_add:
  push p0
  call DUP
  push 42
  push p0
  call CONST
  push p0
  call MOD
  jnz L1
  push p0
  call DUP
  push p0
  call DUP
  push p0
  call ADD
  ret
L1:
  push 42
  push p0
  call CONST
  push p0
  call DUP
  push p0
  call SUB
  ret
    \end{lstlisting}
  \end{subfigure}
  \caption{Compilation steps of the meta-compiler--based threaded code generation.}
  \label{fig:izawa_et_al_tc_flow}
\end{figure}

Meta-compiler--based threaded code generation demonstrates promise as a foundation for multi-tier
\jit{} compilation in a meta-compiler framework, because it simplifies the
integration of a lightweight \jit{} compiler within the \rpython{} framework. However,
meta-compiler--based threaded code generation was
limited to \emph{offline} code generation, and the generated code was not optimized sufficiently. This paper, therefore, proposes optimizations to enable practical code
generation (discussed in Section~\ref{sec:opt_threaded_code}).

\subsection{Simple Object Machine}
\label{sec:som}

Simple Object Machine (SOM) is a dynamically typed language similar to
Smalltalk. It is designed as a subset of Smalltalk for the purpose of teaching
and researching the implementation of a \vm{}~\cite{haupt:2010:sfv:1822090.1822098}.
SOM supports class inheritance, closures, and non-local returns. The SOM family comprises
many \vm{} implementations written in C/C++, Rust, Java, Python, and Smalltalk.

Our proposed two-level SOM is based on PySOM. PySOM is a SOM \vm{} that has AST and
bytecode interpreters written in \rpython{}. We are going to adopt the bytecode
interpreter to implement 2SOM.

\section{Problem}
\label{sec:problem}

While multi-tier \jit{} is used to improve performance in widely used \vm{}s,
implementing a multi-tier \jit{} compilation \vm{} is harder than implementing a single-tier \jit{}
compilation \vm{}. The multi-tier option requires at least the following components of \vm{} developers:

\begin{description}
\item [(Multiple compiler implementations:)] Both lightweight and heavyweight compilers are
required, each generating code at different optimization levels, balancing code generation
time and execution efficiency.
\item [(Profiling mechanism:)] A profiling mechanism is necessary to
  dynamically determine which compiler to invoke. The profiler measures the execution frequency of
  methods or loops in an interpreter or in code generated by a lightweight
  compiler. Profiling data are collected to trigger higher optimization levels for frequently
  executed code.
\item [(Transition mechanism from code to code generated by different compilers:)] A
  mechanism is required to transition between code generated from different compilers,
  enabling seamless transitions between different compilation tiers.
\end{description}

Due to these complexities, it is not trivial to generate a multi-tier \jit{} compilation \vm{} using a
meta-compiler framework. Thus, a na\"ive approach may unnecessarily
increase the implementation costs.
The primary problem lies in the absence of a mechanism for integrating multiple
compilers into such frameworks. One potential approach is to separately implement both
lightweight and heavyweight compilers within the framework, coupled with a profiler to
trigger transitions and manage the control flow between code from different compilers. While
this approach appears feasible, it incurs significant implementation costs comparable to those
already associated with traditional \vm{}s featuring multi-tier \jit{} compilation.

\section{Two-Level Compilation by Integrating Meta-Compiler--Based Threaded Code Generation}
\label{sec:proposal}

To address the problems outlined in Section~\ref{sec:problem}, we propose a novel
methodology for generating a multi-tier \jit{} compilation \vm{} with minimal implementation
costs. This methodology leverages the \rpython{} framework to generate a multi-tier \jit{}
compiler \vm{} directly from interpreter definitions, reducing the need to implement several
necessary components for multi-tier \jit{} compilation from scratch.
As a proof of concept, we generate 2SOM, a two-level multi-tier \jit{}
compiler \vm{} for Simple Object Machine~\cite{haupt:2010:sfv:1822090.1822098},  using \rpython{}.
We will begin by showing the architecture and behavior of 2SOM in
Section~\ref{sec:2som_architecture} and then describe how we realize it in
Section~\ref{sec:2som_tech_detail}.

\subsection{Architecture and Behavior of 2SOM}
\label{sec:2som_architecture}

\begin{figure}[!t]
  \centering
  \includegraphics[width=.9\columnwidth]{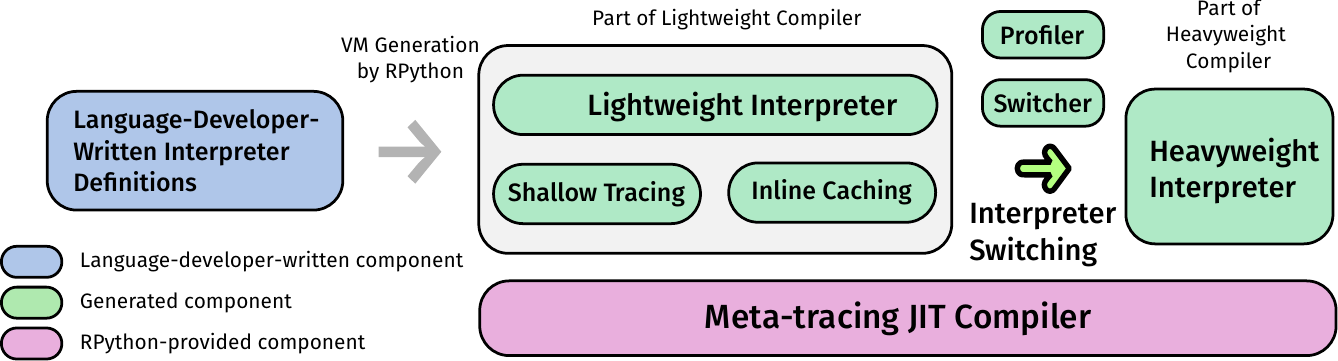}
  \caption{Architecture of 2SOM.}
  \label{fig:overview_integration}
\end{figure}

The architecture of 2SOM, illustrated in Figure~\ref{fig:overview_integration}, uses a
single meta-tracing \jit{} compiler for multi-tier \jit{} compilation to simplify the
implementation of multiple compilers. In this architecture, the interpreter controls
whether lightweight or heavyweight compilation is applied. The interpreter for
lightweight compilation, which is called the \emph{lightweight interpreter} here,
generates threaded code~\cite{izawa:2022:threadedcodegen} in collaboration with the
meta-tracing \jit{} compiler to improve warm-up performance. Conversely, the
interpreter for heavyweight compilation (the \emph{heavyweight interpreter})
generates trace-based, highly optimized code for peak performance. To manage
runtime integration, 2SOM incorporates a profiler and switcher. The profiler identifies
hot spots in code compiled by a lightweight \jit{} compiler, and the switcher transitions execution to
heavyweight compilation when necessary, ensuring efficient optimization and seamless
control flow. The profiler and switcher are implemented as interpreter definitions
for the sake of reducing the need to develop complex components from scratch.

\begin{figure}[!t]
  \centering
  \includegraphics[width=.6\columnwidth]{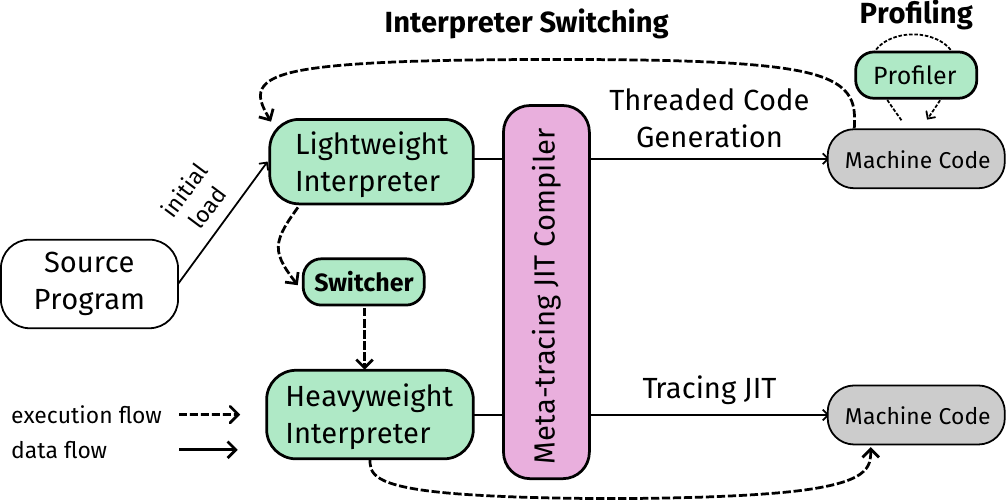}
  \caption{Execution flow of 2SOM.}
  \label{fig:2som_runtime_overview}
\end{figure}

\begin{figure}[!t]
  \centering
  \begin{subfigure}[t]{.475\linewidth}
\begin{lstlisting}[caption={Obtained trace from the lightweight compilation
tier.}, label={lst:ex_trace_lightweight}]
# Trace A
label(strange_add)
call(DUP)
call(CONST, 42)
i2 = call(MOD)
# go to L1 if failed
guard_false(i2) [L1]
# false branch
call(DUP)
call(DUP)
call(ADD)
finish()

# Trace B
# true branch
label(L1)
call(CONST, 42)
call(DUP)
call(SUB)
finish()
\end{lstlisting}
  \end{subfigure}
  \begin{subfigure}[t]{.475\linewidth}
\begin{lstlisting}[caption={Obtained trace from the lightweight compilation
tier.},label={lst:ex_trace_heavyweight}]
# p0 means a red var. in interpreter
label(loop)
i0 = getarrayitem(p0, 0) # get n
i1 = getarrayitem(p0, 1) # get i
i2 = getarrayitem(p0, 2) # get x
# inlining strange_add
i3 = int_mod(i0, 42)    # n % 42
guard_false(i3)         # n % 42 == 0
i4 = int_add(i1, i0)    # ^(n + m)
# inlining finished
i5 = int_add(i2, i4)    # x + strange..
i6 = int_add(i1, 1)     # incr i
i6 = int_le(i6, i0)
guard_true(i6)
setarrayitem(p0, 1, i6) # set to x
setarrayitem(p0, 2, i5) # set to i
jump(loop)
\end{lstlisting}
  \end{subfigure}
  \caption{Obtained traces from 2SOM when executing the program shown in
    Figure~\ref{fig:som_example}.}\label{fig:ex_2level_som_example}
\end{figure}


Next, we illustrate how 2SOM's multi-tier \jit{} compilation works, using the
program shown in Figure~\ref{fig:ex_2level_som_example} and the traces presented
in Figure~\ref{fig:ex_2level_som_example}. An overview of 2SOM's execution flow
is provided in Figure~\ref{fig:2som_runtime_overview}.

As depicted in Figure~\ref{fig:2som_runtime_overview}, the execution of 2SOM
begins in the lightweight interpreter with the \verb|run| function, which serves
as the program's entry point. This function calls \verb|calc:| with an argument
of 10,000. The lightweight interpreter then processes the \verb|calc:| function,
starting with \verb|initialization: x|. Subsequently, the \verb|initialization|
function is compiled into a subroutine-threaded code by the lightweight
compiler. Subroutine-threaded code replaces bytecode instructions with direct
calls to operation handlers, ensuring efficient execution with minimal overhead
compared to heavyweight compilation. The trace obtained from this process is
shown in Listing~\ref{lst:ex_trace_lightweight}.

After initialization, the \verb|calc| function enters a loop that iterates from
1 to \verb|n|. Within this loop, the result of \verb|strange_add: i m: n| is
repeatedly added to \verb|x|. The \verb|strange_add: i m: n| function is initially
compiled by the lightweight compiler. This loop involves frequent backward
jumps, which are monitored by an embedded profiler in the code compiled by the
lightweight interpreter, as depicted by
Figure~\ref{fig:2som_runtime_overview}. Once the frequency of these backward
jumps surpass a predefined threshold, the profiler
identifies the loop as a ``hot'' spot suitable for more aggressive optimization.

At this stage, the interpreter switcher transfers control from the lightweight interpreter
to the heavyweight interpreter. The interpreter switcher recovers the frame used by the lightweight
interpreter and passes it to the heavyweight interpreter for continued execution. The heavyweight interpreter performs
trace-based compilation, applying advanced techniques such as loop unrolling and
inlining. The tracing \jit{} compiler then compiles the entire loop, inlining
and optimizing the \verb|strange_add| function. Hence, the remainder of
the loop executes using highly optimized machine code generated by the
heavyweight compilation tier, thereby improving efficiency while maintaining
correctness. The corresponding trace is presented in Listing~\ref{lst:ex_trace_heavyweight}.

\subsection{Technical Details}
\label{sec:2som_tech_detail}

In this section, we describe the technical details of realizing 2SOM. We describe how each of the three elements essential for realizing two-level \jit{} compilation in 2SOM is achieved, specifically using interpreter definitions and the \rpython{} framework.

\subsubsection{Multiple Compiler Implementations}

To enable multi-tier \jit{} compilation without the complexity of implementing multiple
distinct compilers, we introduce meta-compiler--based threaded code
generation~\cite{izawa:2022:threadedcodegen} into \rpython{}. Consequently,
\rpython{} can generate both generator and compiler from the two interpreter
definitions with different specifications of compiler strategies.
In this setup, the meta-compiler-based threaded code generation works as a lightweight
compiler, while the tracing \jit{} compiler serves as the heavyweight compiler.


Using this technique minimizes the development and maintenance overhead associated with
creating and supporting multiple compilers. Additionally, \vm{} developers can leverage
the interpreter annotations and hints provided by \rpython{} to further customize the
compilation process, tailoring optimization levels to the unique requirements of their
language.

To illustrate this technique, we provide examples of implementing a heavyweight
compiler, which will be followed later by a lightweight compiler example.

\begin{figure}[t]
  \begin{minipage}[t]{.5\columnwidth}
\begin{lstlisting}[language=myPython, caption=Lightweight interpreter for threaded code
  generation.,label=lst:2som_light_interp]
# greens: names of constant
# reds: varying variables (reds)
threadeddriver = JitDriver(
  greens=['bytecodes','pc'],
  reds=['frame'],
  threaded_code_gen=True)

def threaded_interpret(frame):
  # entry point of lightweight compilation
  threadedriver.can_enter_jit(frame=frame,
   bytecodes=frame.bytecodes, pc=frame.pc)

  while True:
    # indicate dispatch loop to
    # the meta-tracing compiler
    threadedriver.jit_merge_point(
      frame=frame,
      bytecodes=frame.bytecodes,
      pc=frame.pc)

    opcode = frame.bytecodes[frame.pc]
    if opcode == ADD:
      handler_add(frame)
    elif opcode == SUB:
      handler_sub(frame)
    # ... other handlers ...
    frame.pc += 1

@enable_threaded_code
def handler_add(frame):
  w_y = frame.pop()
  w_x = frame.pop()
  frame.push(w_x.add(w_y))

# ... other handler functions ...
\end{lstlisting}
  \end{minipage}
  \begin{minipage}[t]{.5\columnwidth}
\begin{lstlisting}[language=myPython, caption=Heavyweight interpreter for tracing
compilation.,label=lst:2som_heavy_interp]
tracingdriver = JitDriver(
  greens=['bytecodes','pc'], reds=['frame'])

def tracing_interpret(frame):
  while True:
    # indicate dispatch loop to
    # meta-tracing compiler
    tracingdriver.jit_merge_point(
      frame=frame,
      bytecodes=frame.bytecodes,
      pc=frame.pc)
    opcode = frame.bytecodes[frame.pc]

    # handle back-ward jump
    if opcode == JUMP_BACKWARD:
      # entry point of heavyweight
      # compilation
      tracingdriver.can_enter_jit(
        frame=frame,
        bytecodes=frame.bytecodes,
        pc=frame.pc)

    elif bytecode == ADD:
      handler_add(frame)
    elif bytecode == SUB:
      handler_sub(frame)
    # ... other handlers ...
    frame.pc += 1

def handler_add(frame):
  w_y = frame.pop()
  w_x = frame.pop()
  frame.push(w_x.add(w_y))

# ... other handler functions ...
\end{lstlisting}
  \end{minipage}
\end{figure}

For the heavyweight compiler, we use the original
interpreter definition originally provided by the \rpython{} framework.
In this interpreter, a \lstinline{jitdriver} is created to configure the constant
variables (greens) and varying variables (reds) in the dispatching loop. Also, hints are
used to guide the \jit{} compiler in optimizing frequently executed code paths. The
hint \lstinline{jit_merge_point} tells the meta-tracing \jit{} compiler the part of
interpreter-dispatching loop, and the \lstinline{can_enter_jit} informs an entry point of
the \jit{} compilation to the meta-tracing \jit{} compiler.

The heavyweight interpreter is defined as displayed by Listing~\ref{lst:2som_heavy_interp}.
The interpreter guides the meta-tracing \jit{} compiler to perform
heavyweight compilation. To indicate the place of the dispatching loop, the
\lstinline{jit_merge_point} hint is placed just after \verb|while True: ..|. To compile
the loop of a source program, \verb|can_enter_jit| hint is placed at the handler for
\verb|JUMP_BACKWARD|, which corresponds to the back-edge jump in a source program.

For the lightweight compiler, we define an interpreter that enables the
meta-tracing \jit{} compiler to perform meta-compiler-based threaded code generation.
The major differences from the heavyweight interpreter are that all handlers are annotated
with \lstinline{enable_threaded_code} and the placement of \lstinline{can_enter_jit}. The
\lstinline{enable_threaded_code}
annotation serves as an instruction to retain calls to handlers in the trace. Moreover, to compile
the method body, \lstinline{can_enter_jit} is placed at the very beginning of the method
call--that is, just before the interpreter is invoked and enters the dispatching loop.






The lightweight interpreter is defined as shown in Listing~\ref{lst:2som_light_interp}.
The handlers \lstinline{handler_add} and \lstinline{handler_sub} are decorated with the
annotation \lstinline{enable_threaded_code}, and the placement of
\lstinline{can_enter_jit} is moved to just before the entry of the dispatching loop.

\subsubsection{Profiling Mechanism}
\label{sec:profiler}

\begin{figure}[t]
  \centering
  \begin{minipage}[t]{.52\columnwidth}
\begin{lstlisting}[language=myPython, caption={Embedded profiler in the interpreter},
label={lst:2som_profiler}]
HOT_THRESHOLD = 1000

def threaded_interpret(frame):
  while True:
    opcode = frame.bytecodes[frame.pc]

    # backward jump should be
    # compiled by heavyweight compiler
    if opcode == JUMP_BACKWARD:
      # increment the counting value
      count = frame.counts.get(frame.pc,0)+1
      frame.counts[frame.pc] = count

      if count == HOT_THRESHOLD:
        # Trigger heavyweight compilation
        raise ContinueInTier2(frame)

    # ... Execute other opcodes ...
    frame.pc += 1
\end{lstlisting}
  \end{minipage}\hfill
\begin{minipage}[t]{.47\columnwidth}
\begin{lstlisting}[language=myPython, caption=Interpreter switcher that dispatches from
  lightweight to heavyweight interpreters, label=lst:2som_switcher]
def interpet_switcher(bytecode):
  frame = Frame(bytecode)
  while True:
    try:
      # Start interpretation first
      # in lightweight interpreter
      w_x = threaded_interpret(frame)
    except ContinueInTier2 as e:
      # Catch thrown ContinueTier2
      # from a generated threaded code,
      # and Continue to execute in
      # heavyweight interpreter
      w_x = tracing_interpret(e.frame)
    return w_x
\end{lstlisting}
\end{minipage}
\end{figure}

Integrating different compilers requires a mechanism to decide when to trigger
heavyweight compilation during executing code generated by a lightweight \jit{} compiler.
We achieve this by generating code fragments that can profile runtime information in the code
compiled by a lightweight \jit{} compiler. These generated profilers are responsible for
monitoring the execution frequency of code spots and triggering recompilation at higher
optimization levels when necessary.

The profiler works using a simple frequency counter that increments each time a
particular code spot, such as a method or loop, is executed. When the counter exceeds a
predefined threshold, the profiler determines that the code spot is ``hot'' and would
benefit from additional optimization. At this point, the profiler triggers the heavyweight
tracing \jit{} compilation for that spot.


Under our methodology, this profiling mechanism provides a flexible method to manage multiple
compilers generated by \rpython{}. \vm{} developers can adjust the thresholds
and profiling strategies to suit the dynamic behavior of their language in their interpreter
definitions.


The example definition of the profiler is depicted in Listing~\ref{lst:2som_profiler}.
This profiler monitors the execution of specific opcodes. In
this definition, backward jumps---typically indicating loops---are identified as key
candidates for heavyweight compilation. When a backward jump opcode
(\lstinline{JUMP_BACKWARD}) is encountered, the profiler increases a counter tied to the
program counter (\lstinline{frame.pc}) stored in the \lstinline{frame.counts} dictionary,
which tracks execution frequencies for each opcode.

If this counter exceeds the predefined \lstinline{HOT_THRESHOLD} (set to 1000 in this
case), the profiler marks the code spot as ``hot'' and suitable for further
optimization. At this point, the profiler triggers heavyweight compilation by using a
global jump. Because \rpython{} does not provide global jumping, we use an
exception mechanism instead. This exception, called \lstinline{ContinueTier2},  halts the
lightweight interpreter and switches control to the heavyweight interpreter. To make the
transition work correctly, the exception contains an execution context. The exception is
used in the transition mechanism described in Section~\ref{sec:transition}.


\subsubsection{Transition Mechanism Between Generated Code}
\label{sec:transition}

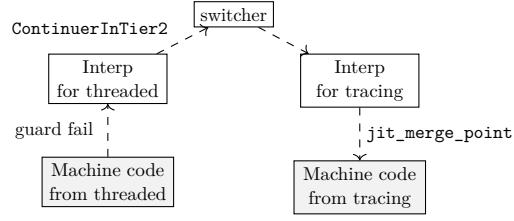
\begin{figure}[t]
  \centering
  \begin{tikzpicture}[auto,transform shape,scale=0.7,align=center]
    \node at (0,0) [draw,rectangle] (switcher) {switcher};
    \node [draw,below left=5mm and 5mm of switcher,text width=20mm]
    (tier1) {Interp\\for threaded};
    \node [draw,below right=5mm and 5mm of switcher,text width=20mm,align=center]
    (tier2) {Interp\\for tracing};
    \node [draw,fill=gray!10,below=10mm of tier2,text width=22.5mm] (native2)
    {Machine code\\from tracing};
    \node [draw,fill=gray!10,below=10mm of tier1,text width=22.5mm] (native1)
    {Machine code\\from threaded};

    \path (native1) edge [dashed,->] node [midway,xshift=-1ex] {guard fail} (tier1);
    \path (tier1) edge [dashed,->] node [midway,xshift=-1ex] {\texttt{ContinuerInTier2}} (switcher);
    \path (switcher) edge [dashed,->] (tier2);
    \path (tier2) edge [dashed,->] node {\texttt{jit\_merge\_point}} (native2);
  \end{tikzpicture}
  \caption{Overview of interpreter switching technique.}
  \label{fig:interp_switching}
\end{figure}



Finally, we need a mechanism to effect transitions from code generated by a lightweight \jit{}
compiler to that generated by a heavyweight \jit{} compiler.
This is achieved through the \emph{interpreter switching} technique, which redirects the
execution in code compiled by the lightweight compiler to the code compiled by the
heavyweight compiler.

We overview the technique in Figure~\ref{fig:interp_switching}.
When the profiler, the function of which is explained in Section~\ref{sec:profiler}, detects
a hot spot suitable for tracing \jit{} compilation, a guard fail occurs. Then, a global jump
from code generated by a lightweight \jit{} compiler to the lightweight interpreter happens.
Next, the lightweight interpreter
throws an exception to move to the interpreter switcher. The switcher catches it and
resumes the execution context. Next, it launches the heavyweight interpreter with the
resumed execution context. If the running program is already compiled by the heavyweight
compiler, the control goes to the compiled machine code via
\verb|jit_merge_point|. Otherwise, the program is interpreted and will be compiled via
\verb|can_enter_jit| in the heavyweight interpreter.

The interpreter switcher is implemented as described in Listing~\ref{lst:2som_switcher}.
The \lstinline{interpret_switcher} function handles this transition. It starts with the
lightweight interpreter and, upon catching a \lstinline{ContinueTier2} exception thrown
from \lstinline{threaded_interpret}, passes
the execution context (\lstinline{e.frame}) to the heavyweight interpreter
\lstinline{tracing_interpret}. Both
interpreters return a result, \lstinline{w_x}, to the caller, ensuring consistent
execution flow.

Together with the profiling and interpreter switching technique explained in
sections~\ref{sec:profiler} and~\ref{sec:transition}, we can maintain
correctness while optimizing performance. By leveraging the interpreter for managing
transitions, the framework provides a basis for multi-tier \jit{} compilation.

However, this mechanism can introduce significant runtime overhead because the control bypasses the interpreter when it goes to other code. We evaluate the overhead by measuring the peak performance of 2SOM's two-level compilation in section~\ref{sec:evaluation}.

\section{Improving Meta-Compiler--Based Threaded Code Generation}
\label{sec:opt_threaded_code}

In 2SOM, we use meta-compiler--based threaded code
generation~\cite{izawa:2022:threadedcodegen} for lightweight compilation. However,
its integration with a meta-tracing \jit{} compiler in 2SOM has exposed
problems, as the earlier implementation~\cite{izawa:2022:threadedcodegen} focused solely
on code generation.
Therefore, we analyzed the technical problems that arise when introducing \tcg{} to the meta-tracing compiler framework and introduced a new technique and optimization to enhance its efficiency and compatibility within the meta-tracing \jit{} compiler.

\subsection{Runtime Problem of the Meta-Compiler-Based Threaded Code Generation}

\begin{figure}[!t]
\begin{minipage}[t]{.48\linewidth}
\begin{lstlisting}[language=Smalltalk,caption={Example of SOM program that has a
side effect.},label={lst:strange_sum_arr}]
"Compute the sum of arr"
strange_sum_arr: arr index: i
  sum: n = (
  [ i <= arr length ]
    ifTrue: [
     sum_array: arr
      index: (i + 1)
      sum: (n + (arr at: i)).
    ]
    ifFalse: [
     "Clear all the elements"
     arr clear. ^ n  ] )

"Entry point"
run = (
  | arr |
  "Create an array where"
  "all elements are 1"
  arr :=
    Array new: 30 putAll: 1.
  strage_sum_arr: arr
    index: 1 sum: 0. )
\end{lstlisting}
\end{minipage}\hfill
\begin{minipage}[t]{.48\linewidth}
  \centering
  \begin{lstlisting}[caption=Simplified bytecode from
Listing~\ref{lst:strange_sum_arr}.,label={lst:shallow_tracing_source}]
strange_sum_arr:
 DUP0 # arr
      # arr length
 CALL(length, frame)
 DUP2 # i
 LE   # i <= arr length
 JUMP_IF_FALSE L1
 DUP1 # i
 CONST 1
 ADD  # i + 1
 DUP1 # arr
 CALL(at, frame)
      # arr at: i
 DUP2 # n
 ADD  # n + (arr at: i)
 JUMP strange_sum_arr
L1:
 DUP0 # arr
 CALL(clear, frame)
 DUP2 # n
 RET
\end{lstlisting}
\end{minipage}\hfill
\begin{minipage}[t]{\columnwidth}
  \begin{multicols}{2}
\begin{lstlisting}[language=myPython,caption={Handler function that interprets \texttt{CALL} instruction.},
label={lst:handler_call}]
def threaded_interpret(frame):
  ...
  while True:
    ...
    if opcode == CALL:
      handler_CALL(frame)
    # ... XXX ...

@dont_look_inside
def handler_CALL(frame):
  method = frame.bytecodes[frame.pc+1]
  newframe = crate_frame(frame, method)
  w_x = threaded_interpret(newframe)
  frame.push(w_x)
\end{lstlisting}
  \end{multicols}
\end{minipage}
\end{figure}

Meta-compiler--based threaded code generation exhibits unsatisfactory
runtime execution and performance.
In particular, the following technical problems occur when we use it in 2SOM.

\begin{enumerate}
\item If the definition of the lightweight interpreter has side effects, tracing the
  lightweight interpreter leads to an incorrect state due to tracing both
  conditional branches in a row.
\item Function calls always become indirect in the compiled code.
\end{enumerate}

The first problem occurs because meta-compiler--based threaded code generation and tracing
\jit{} compilation use different compilation scopes (method-based for the threaded code
generation and trace-based for tracing compilation) but rely on the same compilation engine. For
example, when performing threaded code generation, as described in the previous section,
branches in the code are linearized into a single sequence during tracing. This means that
even branches that would not be executed during actual program runs are included in the
trace. If these unnecessary branches are executed during tracing, it can lead to an
invalid runtime state.

Consider a simple \verb|if-then-else| statement where the \verb|then| branch is
always taken during execution. During threaded code generation, both branches
are traced at compilation time as described in
Section~\ref{sec:background}. Although the meta-tracing \jit{} compiler executes
the source program with the interpreter at compilation, tracing the \verb|else|
branch can modify variables or states, causing inconsistencies. For example, the
program in Listing~\ref{lst:strange_sum_arr}\footnote{Its simplified bytecode
  appears in Listing~\ref{lst:shallow_tracing_source}.} computes the sum of an
array: the \verb|then| clause (receiver of \verb|ifTrue|) adds each value to
\verb|n|, while the \verb|else| clause (receiver of \verb|ifFalse|) returns
\verb|n|. However, executing the \verb|clear| function in the \verb|else|
clause, which sets the array to zeros, corrupts the runtime state if na"ively
traced by the meta-compiler--based threaded code generation.

The second problem arises because the meta-tracing \jit{} compiler could not identify the target
of a handler function wrapped in a call instruction during tracing, when the handler
represented a method invocation.
In general, calling a dynamically bound method takes longer time than calling a
statically bound method. This is because a system needs to look up a correct method by using the
runtime class of a callee method. The same kind of problem occurs in threaded code
generation.

For example, consider a handler function \verb|handler_CALL| that dynamically dispatches a
method call as illustrated in Listing~\ref{lst:handler_call}. During tracing, the meta-tracing
\jit{} compiler encounters the call instruction for \verb|handler_CALL| but cannot identify the
callee method compiled at tracing. Without knowing whether the
target method is compiled, the compiler cannot optimize the method call, which
leads to worse performance.

In \tcg{}, this could result in repeated interpretation or
inefficient indirect calls, even if the invoked method is a commonly used one that would
benefit from inlining or specialized optimizations. This lack of optimization is
particularly problematic for workloads with frequent method invocations, as it leads to
missed performance improvements.

\subsection{Our Solutions}

In this section, we show the solutions, namely, shallow tracing and direct calls by inline
caching to the first and second problems.

\subsubsection{Shallow Tracing}

\begin{figure}[!t]
  \centering
  \begin{subfigure}[t]{.48\columnwidth}
    \centering
    \includegraphics[width=\linewidth]{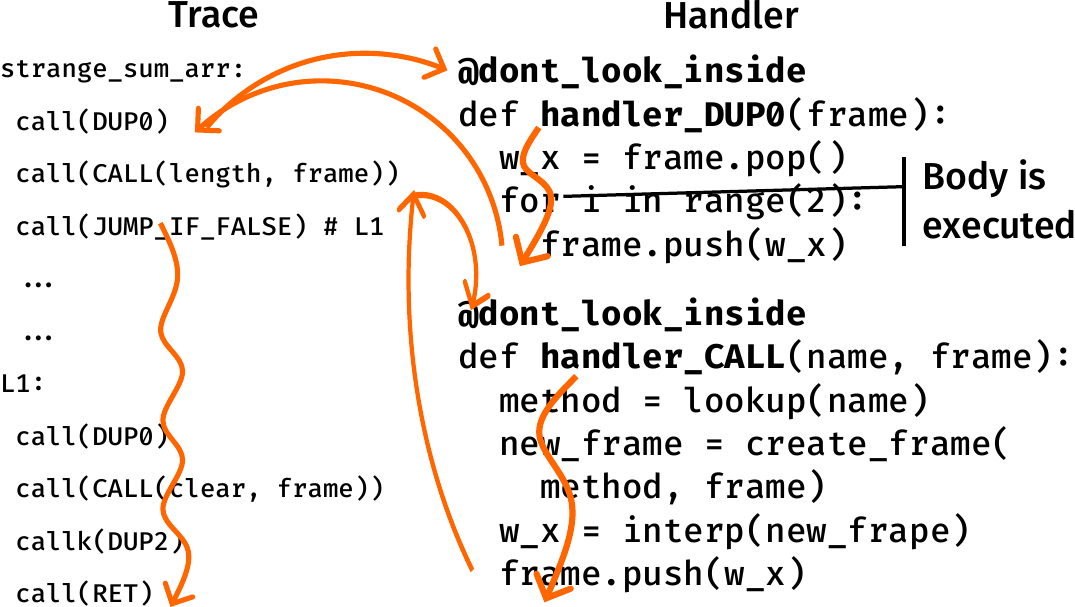}
    \caption{Flow of tracing before applying shallow tracing.
      The left-hand side and right-hand side show the trace and the handler
      functions, respectively.}
  \end{subfigure}\hfill
  \begin{subfigure}[t]{.48\columnwidth}
    \centering
    \includegraphics[width=\linewidth]{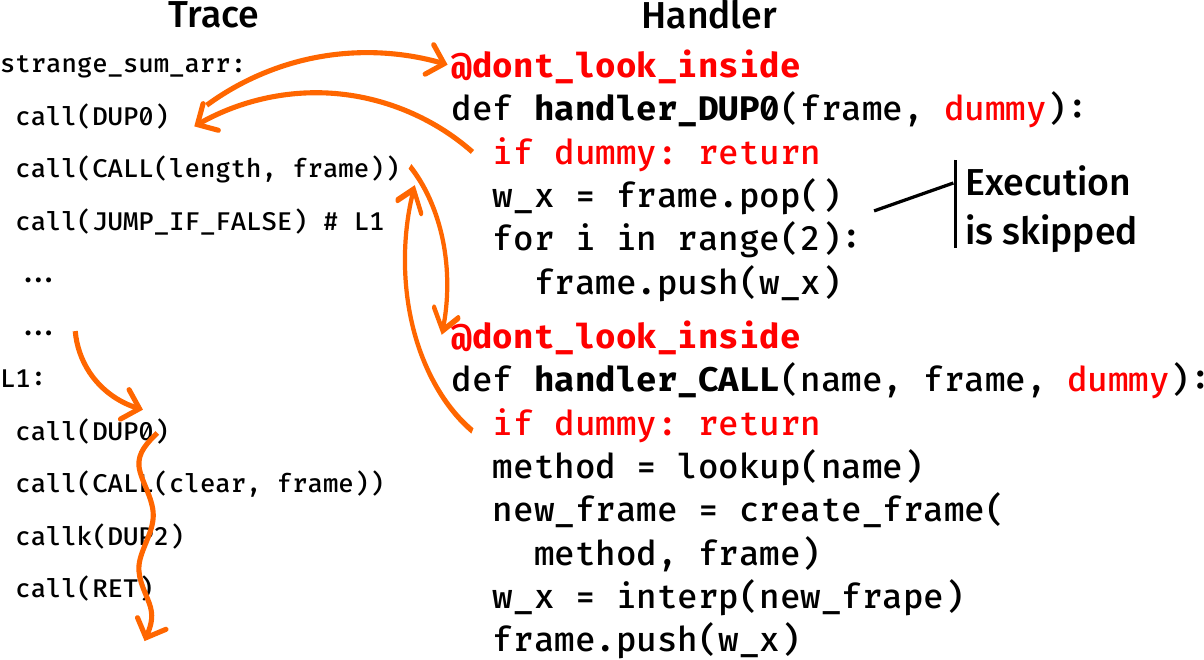}
    \caption{Flow of tracing after applying shallow tracing.
      The left-hand side and right-hand side show the trace and the handler functions,
      respectively.}
  \end{subfigure}
  \caption{The problem with na\"ively integrating meta-compiler-based threaded code
    generation with a meta-tracing \jit{} compiler (left-hand side), and our
    solution to it (right-hand-side). In each side,
    orange lines are the flow of tracing. In na\"ive integration,
    the body of the handler function is executed during tracing, which can lead
    to side effects that corrupt the interpreter's state. To resolve this, we
    introduce shallow tracing, a technique that prevents the execution of the
    handler function's body during tracing, thereby avoiding such side effects.}
  \label{fig:exec_flow_shallow_tracing}
\end{figure}

\begin{figure}[t]
\begin{minipage}[t]{.5\columnwidth}
\begin{lstlisting}[language=myPython,
caption={Na\"ive version of an interpreter that supports shallow tracing.},
label={lst:shallow_tracing}]
def threaded_interpret(frame):
  while True:
    opcode = frame.bytecodes[frame.pc]
    if opcode == ADD:
      # returns true while tracing and
      # executing machine code
      if we_are_jitted():
        # set 'dummy' True
        # to avoid side-effects
        handler_add(frame, dummy=True)
      else:
        # set 'dummy' False
        # execute body
        handler_add(frame, dummy=False)

    elif opcode == SUB:
      if we_are_jitted():
        handler_sub(frame, dummy=True)
      else:
        handler_sub(frame, dummy=False)
    # ... other handlers ...
    frame.pc += 1

# leaves only handler call but
# executes the body
@dont_look_inside
def handler_add(frame, dummy):
  # immediately return when dummy
  # is true to avoid side-effects
  if dummy: return
  w_y, w_x = frame.pop(), frame.pop()
  frame.push(w_x.add(w_y))

@dont_look_inside
def handler_sub(frame, dummy):
  if dummy: return
  # ... subtract two elements in
  # frame.stack ...
\end{lstlisting}
\end{minipage}
\begin{minipage}[t]{.5\columnwidth}
\begin{lstlisting}[escapeinside={!}{!},
caption={Na\"ive version of a generated trace by shallow tracing.},
label={lst:naive_shallow_tracing_trace}]
label(strange_sum_arr)
call(handler_DUP, p0, True)
call(handler_CALL, "length", p0, True)
call(handler_DUP2, p0, True)
i1 = call(handler_LE, p0, True)
# i1 is initialized with a default value (0)
guard_false(i1)
call(handler_DUP1, p0, True)
call(handler_CONST, p0, 1, True)
call(handler_ADD, p0, True)
call(handler_ADD, p0, "at", True)
call(handler_DUP2, p0, True)
call(handler_ADD, p0, True)
jump(strange_sum_arr)

label(L1)
call(handler_DUP0, p0, True)
call(handler_CALL, "clear", p0, True)
finish()
\end{lstlisting}
\end{minipage}
\end{figure}

For the first problem, we propose the \emph{shallow tracing} technique.
Shallow tracing prevents side effects during tracing by introducing lightweight
annotations within the interpreter. These annotations guide the tracing process to leave
only call instructions to handler functions, while the actual execution of these handlers
is avoided. In contrast, during the interpretation and execution of generated machine code,
the annotations have no effect, ensuring seamless runtime execution.

The fundamental reason for needing shallow tracing lies in the fact that, while
the \verb|dont_look_inside| annotation retains the call instruction to the
handler function, it still executes the function's body, using the
interpreter's state. This
is not a problem in standard tracing compilation. However, in method-based
threaded code generation, both branches of a conditional are traced. If side
effects occur during the tracing of one branch, they propagate to the tracing of
the other branch, corrupting the interpreter's state.

Shallow tracing suppresses the execution of the body that would otherwise occur
under \verb|dont_look_inside|.  This technique relies on the following two key
annotations and the dummy flag mechanism:

\begin{itemize}
\item \verb|we_are_jitted|: Checks whether the execution is currently within a traced
  and compiled context, and
\item \verb|dont_look_inside|: Ensures that only call instructions to annotated functions
  are traced without inlining their bodies of annotated functions, and
\item \verb|dummy| flag mechanism: the dummy flag is inserted into every handler.
It is turned into \verb|True| inside the then branch conditioned with \verb|we_are_jitted|.
When this flag is on, the dummy return is triggered. When this trace is compiled and executed,
the dummy flag is turned into \verb|False|,  allowing the handler's body to execute correctly.
\end{itemize}

Its overall execution flows are compared in Figure~\ref{fig:exec_flow_shallow_tracing}: the left-hand
side illustrates the trace without shallow tracing, while the right-hand side shows
the execution flow with shallow tracing applied. Shallow tracing introduces a
\emph{dummy flag} and a \emph{dummy return} for each handler (except for
control-flow--related handlers like \verb|JUMP_IF_FALSE|). Also, the
implementation in an interpreter and the resulting trace are shown in
Listings~\ref{lst:shallow_tracing} and~\ref{lst:naive_shallow_tracing_trace},
respectively.

\begin{figure}[!t]
  \centering
  \begin{minipage}[t]{.48\columnwidth}
\begin{lstlisting}[language=myPython,
caption=Annotated version of the interpreter design for shallow tracing.,
label=lst:shallow_tracing_annot]
def threaded_interpret(frame):
  while True:
    opcode = frame.bytecodes[frame.pc]
    if opcode == ADD:
      handler_add(frame)
    elif opcode == SUB:
      handler_sub(frame)
    frame.pc += 1

@enable_threaded_code
def handler_ADD(frame):
  w_y = frame.pop()
  w_x = frame.pop()
  frame.push(w_x.add(w_y))

@enable_threaded_code
def handler_SUB(frame):
  # subtract two elements in frame.stack
\end{lstlisting}
\end{minipage}\hfill
\begin{minipage}[t]{.48\columnwidth}
\begin{lstlisting}[language=myPython,
caption=Generated stub and original handlers.,
label=lst:shallow_tracing_gen_stub]
def threaded_interpret(frame):
  while True:
    opcode = frame.bytecodes[frame.pc]
    if opcode == ADD:
      if we_are_jitted():
        stub_ADD(frame, True)
      else:
        ADD(frame)
    elif opcode == SUB:
      if we_are_jitted():
        stub_SUB(frame, True)
      else:
        SUB(frame)
    frame.pc += 1

@dont_look_inside
def stub_ADD(frame, dummy):
  if dummy: return
  ADD(frame)

@dont_look_inside
def stub_SUB(frame, dummy):
  if dummy: return
  SUB(frame)

def ADD(frame):
  # ... add two elements ...

def SUB(frame):
  # ... subtract two elements ...
\end{lstlisting}
  \end{minipage}
  \begin{minipage}[t]{.49\linewidth}
  \centering
\begin{lstlisting}[caption={Before applying the handler optimization in shallow
tracing.},label={lst:shallow_tracing_befor_opt},
escapeinside={!}{!}]
label(strange_sum_arr)
call(stub_DUP, p0, `True`)
call(stub_CALL, "length", p0, `True`)
call(stub_DUP2, p0, `True`)
i1 = call(stub_LE, p0, `True`)
# i1 is initialized with a default value (0)
guard_false(i1)
call(stub_DUP1, p0, `True`)
call(stub_CONST, p0, 1, `True`)
call(stub_ADD, p0, `True`)
call(stub_CALL, p0, "at", `True`)
call(stub_DUP2, p0, `True`)
call(stub_ADD, p0, `True`)
jump(strange_sum_arr)

label(L1)
call(handler_DUP0, p0, `True`)
call(handler_CALL, "clear", p0, `True`)
finish()
\end{lstlisting}
  \end{minipage}\hfill
  \begin{minipage}[t]{.49\linewidth}
\begin{lstlisting}[
caption={After applying the handler optimization in shallow tracing.},
label={lst:shallow_tracing_after_opt},
escapeinside={!}{!}]
label(strange_sum_arr)
call(DUP, p0)
call(CALL, "length", p0)
call(DUP2, p0)
i1 = call(LE, p0)
# i1 is initialized with a default value (0)
guard_false(i1)
call(DUP1, p0)
call(CONST, p0, 1)
call(ADD, p0)
call(CALL, p0, "at")
call(DUP2, p0)
call(ADD, p0)
jump(strange_sum_arr)

label(L1)
call(handler_DUP0, p0)
call(handler_CALL, "clear", p0)
finish()
\end{lstlisting}

  \end{minipage}
\end{figure}

\paragraph{Optimizing Shallow Tracing}

Manually implementing the code in Listing~\ref{lst:shallow_tracing} is complex
and time-consuming for \vm{} developers. Na"ive insertion of a dummy flag into
handlers also risks performance overhead. To address this, we automate the
interpreter generation at \vm{} generation time, while optimizing
shallow-tracing code by eliminating redundant \verb|dummy| flag checks.

We introduce a new annotation, \verb|enable_threaded_code|, to automate this
process. Annotated handlers are automatically marked with
\verb|dont_look_inside| and a dummy flag, and their call sites are wrapped with
\verb|we_are_jitted|, leveraging the \rpython{} \vm{} generation
process~\cite{rigo:2006:10.1145/1176617.1176753}. Developers only need to
annotate each handler function, as shown in
Listing~\ref{lst:shallow_tracing_annot}, and the annotated interpreter is
transformed into the structure in Listing~\ref{lst:shallow_tracing_gen_stub}.

To mitigate the dummy flag overhead, we apply a two-step optimization. First,
for each annotated handler, we generate both a stub handler (with a dummy check)
and a real handler (without the check) under the same name. Second, during
shallow tracing, only stub handlers are traced to build the
instruction-to-handler mapping; after tracing, they are replaced with real
handlers at compile time. Listings~\ref{lst:shallow_tracing_befor_opt}
and~\ref{lst:shallow_tracing_after_opt} show the traces before and after this
optimization.

\subsubsection{Direct Calls with Inline Caching}

Inline caching~\cite{deutsh:1984} is a well-known runtime optimization technique used to
accelerate method dispatch in object-oriented programming languages such as Smalltalk-80,
SELF, and Java. It optimizes method calls by caching method lookups directly at the call
site, avoiding redundant type resolution when the same types are repeatedly
encountered. If the cached type matches the runtime type of the receiver, a fast path is
taken; otherwise, the system falls back to a slow path for dynamic method
resolution~\cite{hozle:1991:10.1007/bfb0057013}.

\begin{figure}[t]
  \begin{minipage}[h]{.48\linewidth}
    \centering
    \includegraphics[width=\linewidth]{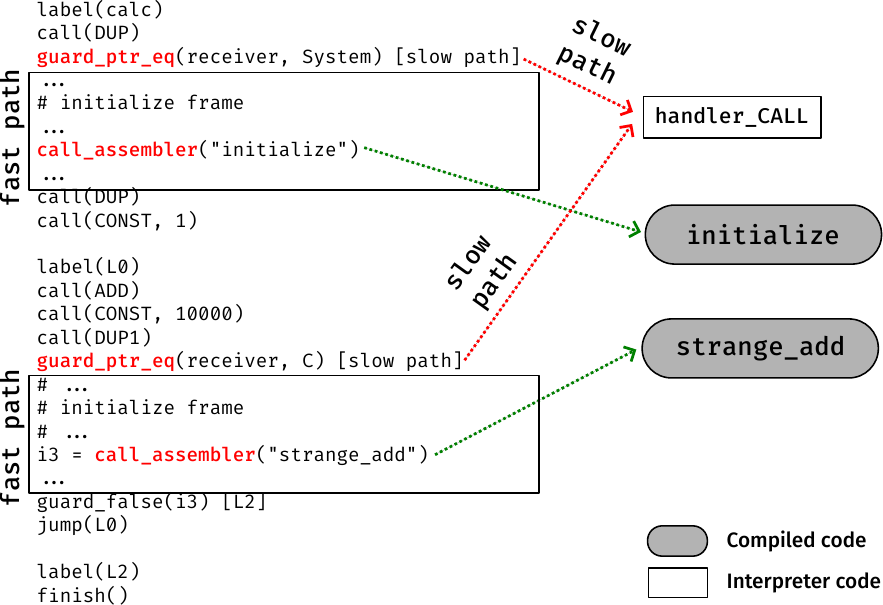}
    \caption{Overview of direct calls with inline caching.}
    \label{fig:overview_ic}
  \end{minipage}\hfill
  \begin{minipage}[h]{.48\linewidth}
    \begin{lstlisting}[
language=myPython,
caption={Interpreter definition instrumented for inline caching},
label={lst:interp_dispatch_handler_call_ic},
escapechar=`]
def threaded_interpret(frame):
  while True:
    jit_merge_point(..)
    ...
    if opcode == CALL:
      # check whether the runtime type of
      # this method equals to the
      # recorded runtime type
      method = frame.stack[frame.pc]
      if `\textcolor{red}{\texttt{check\_type}}`(method, frame.pc):
        # if it matches, calling the
        # method with call_assembler
        `\textcolor{red}{\texttt{call\_assembler}}`(interp(frame))
      else:
        # otherwise, call handler_CALL
        handler_CALL(frame)

def handler_CALL(frame):
  method = pop(stack)
  if not we_are_jitted():
     # during interpreting (not jitted),
     # record runtime method types at pc
     `\textcolor{red}{\bf\texttt{record\_type}}`(pc, method.type)
  r = threaded_interpret(frame)
  frame.push(r)
\end{lstlisting}
  \end{minipage}
\end{figure}

In the absence of inline caching in \tcg{}, method calls to compiled code are
executed as indirect calls. Such calls bypass \verb|jit_merge_point|, which can lead to
significant performance degradation when methods are invoked repeatedly. Direct calls with
inline caching alleviate this by converting indirect calls into direct calls, thereby
improving execution speed. Figure~\ref{fig:overview_ic} provides an overview of this
transformation at runtime, revealing the transition from indirect to direct calls enabled by
inline caching.

To integrate inline caching efficiently, we leverage the annotation mechanism of the
\rpython{} framework instead of implementing it from scratch. Inline caching is realized
by instrumenting the interpreter with runtime type collection, validation, and direct call
conversion. During interpretation, runtime type information---such as the receiver's type
and the associated method---is collected. This collected type information is then validated
at runtime, enabling the conversion of indirect calls into direct calls.

At runtime, our inline caching implementation works as follows:

\begin{itemize}
\item Fast Path: If the runtime type matches the cached type, the system directly calls
  the compiled method using the \verb|call_assembler| instruction.
\item Slow Path: If the type validation fails, the system falls back to the
  \verb|handler_CALL| function, which dynamically resolves the method and records type
  information for future optimizations.
\end{itemize}

The implementation uses the \rpython{} annotation mechanism. Two annotations
were originally developed:

\begin{itemize}
\item \verb|record_type|: records the runtime type of a method during interpretation.
\item \verb|check_type|: validates the runtime type against the cached type.
\item \verb|call_assembler|: performs the direct call to the compiled method.
\end{itemize}

Listing~\ref{lst:interp_dispatch_handler_call_ic} illustrates how these annotations are
used in the interpreter:

Based on the implementation, the compilation with direct calls with inline caching works
as follows:

\begin{enumerate}
\item During interpretation, the \verb|handler_CALL| function records the runtime type of the
  receiver using the \verb|record_type| annotation.
\item During tracing, the \verb|CALL| handler, \verb|check_type| validates the runtime type
  of the receiver. If the validation succeeds, the \verb|call_assembler| instruction is
  invoked to perform a direct call to the compiled method.
\item During compilation, \verb|check_type| is transformed into a guard instruction that
  ensures the receiver type matches the recorded type. Similarly, \verb|call_assembler| is
  converted directly into an assembler call instruction in the resulting trace.
\end{enumerate}

\section{Evaluation}
\label{sec:evaluation}



In this section, we address the three questions introduced at the beginning of
this paper by evaluating the performance of the two-level compilation implemented in
2SOM. We begin by describing the benchmarking programs and the execution environment used in
the evaluation. Each research question is then addressed by analyzing both the
warmup performance and peak performance of two-level compilation and meta-compiler--based
threaded code generation using 2SOM. Since the measurement methods differ for each case, they are described in separate sections of their own.

\subsection{Methodology}
\label{sec:evaluation_setup}

Throughout this evaluation, we basically use the SOM benchmarks~\cite{sombench}, an
extended version of the Are We Fast Yet benchmark~\cite{10.1145/2989225.2989232}, which
includes programs such as PageRank and graph search algorithms.

To answer Q1 and Q2 in particular, we should analyze
the performance of two-level \jit{} using real-world large-scale programs. However,
SOM benchmark suite does not have such programs. Thus, we synthesize a program that has
workloads similar to real-world large-scale programs.

First, we examined the relationship when methods in the DaCapo benchmark~\cite{blackburn:2006:10.1145/1167473.1167488}
were sorted in descending order based on the number of method calls. Using the least squares method
to analyze this relationship, we found that the $R^2$ value exceeded 0.98, indicating
a high correlation. Therefore, we aimed to approximate the ranking correlation between
DaCapo’s method call counts and the method order using the SOM benchmark suite.

We generated 20 variants of \emph{synthesized benchmark program} using the following procedure:
(1) place each program SOM's benchmark suite in order, (2) manually tune the number of internal iterations
for each program so that the correlation between DaCapo’s method call counts and method ranks is as close as possible.
(3) Finally, from the program created in steps (1) and (2), generate the remaining 19
variants by randomly shuffling the program execution order.%
\ifextendedversion%
  The details are described and discussed in Appendix~\ref{app:construct_large_workload_app}.
\else%
  The details are described and discussed in Appendix A of the extended
  version~\cite{ecoop2025extended}.
\fi%

In Q1, we measure the elapsed time obtained from the first iteration of each program and repeat this
measurement 2,000 times. We calculate their medians, averages, and variances.



We evaluate the warm-up and peak performances against six execution modes:

\begin{itemize}
\item Interpreter-only execution: execution using 2SOM's
  interpreter without any form of compilation.
\item Threaded code generation: meta-compiler--based threaded code generation implemented
  in 2SOM, offering a lightweight approach to improving execution efficiency.
\item Tracing \jit{}: execution utilizing \rpython{}'s meta-tracing compiler.
\item Tracing \jit{} with high threshold: A variation of the tracing compilation
  where the compilation threshold is increased by a factor of three. This configuration is
  designed to investigate the fundamental differences between a higher-threshold tracing
  and two-level \jit{} compilations.
\item Two-level \jit{}: A combined strategy that combines both threaded code
  generation and tracing compilation, leveraging the strengths of both techniques for
  improved execution performance.
\item TrufleSOM: execution using TruffleSOM, which is another SOM built with the
    Truffle framework~\cite{wurthinger:2012}, to compare the performance of 2SOM with
    a different language implementation. Note that multi-tier \jit{} compilation~\cite{prokopec:2021:graalmulti}
    is enabled.
\end{itemize}

In Q2, we evaluate the peak performance of 2SOM. We use
the synthesized programs used in Q1 in addition to the SOM microbenchmark programs.
Each program is executed 2,000 times, and the average elapsed times are
calculated based on the last half of the iterations to omit the effect of \jit{} compilation.
We compare the following execution models: tracing \jit{},
tracing \jit{} with a higher threshold, and two-level \jit{}.

In Q3 (performance of threaded code), we assess the improvements introduced by \tcg{} by utilizing
the SOM microbenchmarks. Specifically, we measure the peak performance of these benchmarks to
evaluate the effectiveness of \tcg{} enhancements.


The 2SOM codebase is available at Zenodo~\cite{2som:2025}.
To build 2SOM, we use GCC version 14.2.1 and our customized PyPy, which is
available at Zenodo~\cite{pypy_threaded:2025}. For measuring time, we use
\verb|monotonic_clock| by our RPython extension~\cite{rtime_ext:2025}.
For TruffleSOM, we use revision e9d8032 of the repository hosted on GitHub\footnote{
\url{https://github.com/SOM-st/TruffleSOM}}, along with GraalVM Community Edition
version 23.0.2.

All evaluations are conducted on a system running Ubuntu 22.04.1 LTS, equipped with a quad-core
Intel Core i7-6700 CPU and 32 GB of RAM. To ensure reliable and consistent measurements, we use
ReBench~\cite{rebench:2018}, a benchmarking framework designed to manage the evaluation process
and minimize measurement noise.

\subsection{Q1: Does Two-Level Compilation Improve Warm-Up Performance?}

\begin{figure}[t]
  \centering
  \includegraphics[width=\columnwidth]{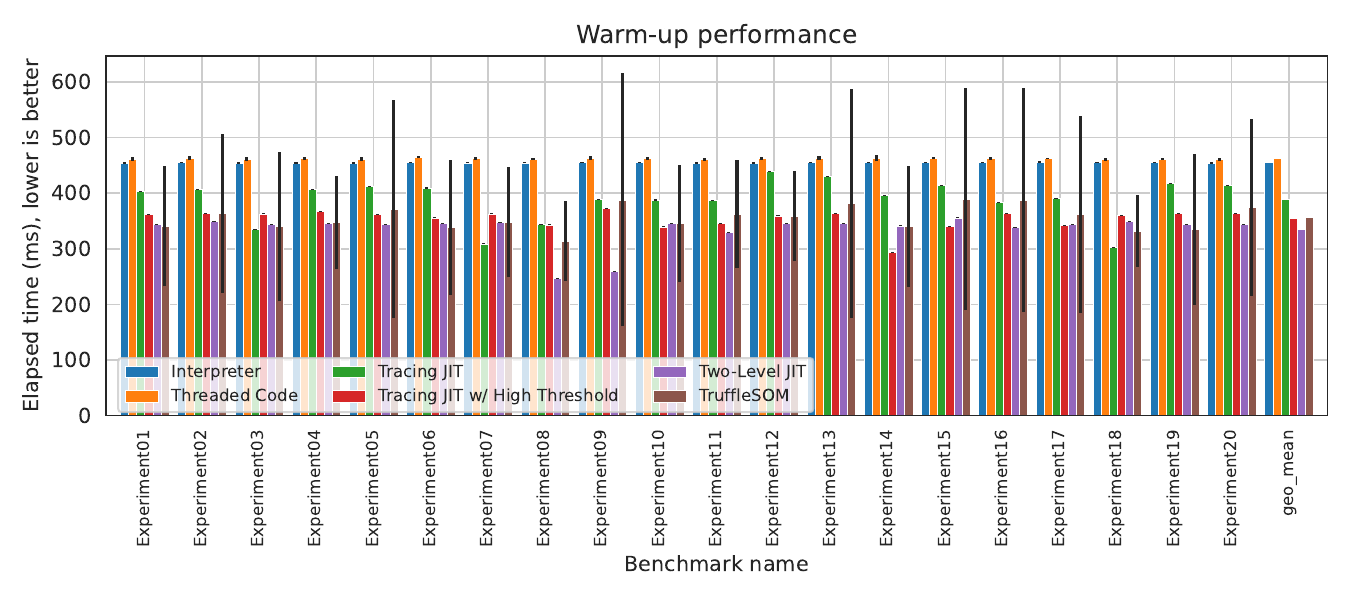}
  \caption{Warm-up performance (execution time of the first iteration)}
  \label{fig:rq1_result}
\end{figure}

In this section, we answer Q1 by evaluating the warm-up performance of two-level
\jit{} in 2SOM. The results of these measurements are presented in
Figure~\ref{fig:rq1_result}, which shows the elapsed times with error bars displaying their variances.



As shown in Figure~\ref{fig:rq1_result},
two-level compilation demonstrated the best warm-up performance; it executed approximately
15\% faster than tracing \jit{} alone. Furthermore, even when
compared to tracing compilation with a threshold increased by a factor of three and TruffleSOM,
two-level compilation outperformed by approximately 5\%.

\begin{figure}[t]
  \centering
  \includegraphics[width=\linewidth]{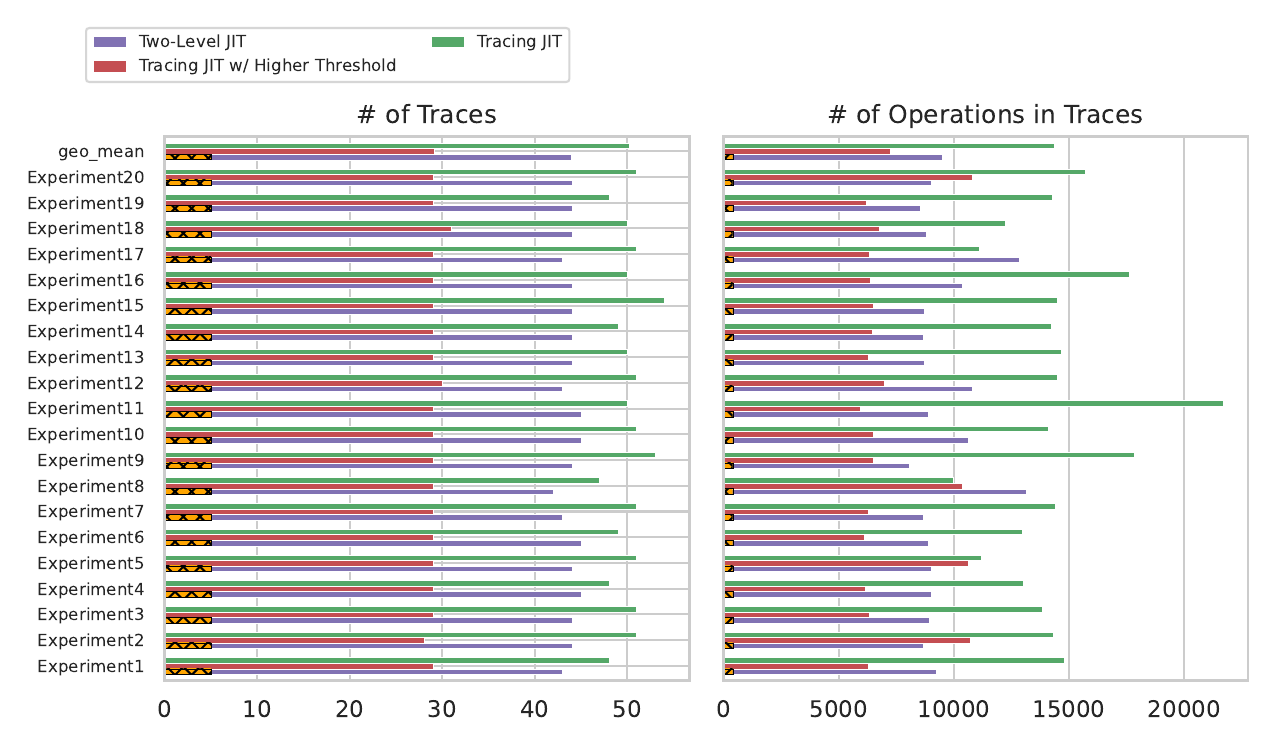}
  \caption{Number of operations and traces in the following strategies: two-level \jit{},
    tracing \jit{} with high threshold, and tracing \jit{} in the evaluation of warm-up
    performance. The orange stitched bar inside the two-level \jit{} shows the number of
    operations and traces generated from the threaded code generation.}
  \label{fig:rq1_num_ops_trace}
\end{figure}

To statistically validate whether there were differences in the
data for each pair---two-level and tracing \jit{}s, as well as two-level
\jit{} and tracing \jit{} with a high threshold---we conducted the Wilcoxon
signed-rank test. The results revealed that for the two-level and tracing
\jit{}s pair, $p = 0.0314\%$, and for the two-level \jit{} and tracing
\jit{}  with a higher threshold pair, $p = 2.151\%$. In both cases, the null
hypothesis was rejected.

These findings indicate that two-level \jit{} improves warm-up performance compared
to \rpython{}'s single-tier compilation. Furthermore, they suggest that two-level
\jit{} enhances warm-up performance more effectively than optimizing compilation
strategies by limiting the compilation target scope, such as adjusting thresholds to
reduce compilation time.

In addition, we break down the traces generated from these strategies: two-level
\jit{}, tracing \jit{} with high threshold, and tracing \jit{}. The results are shown in
Figure~\ref{fig:rq1_num_ops_trace}.

The first breakdown (the left-hand side of Figure~\ref{fig:rq1_num_ops_trace}: number of
traces) shows that two-level \jit{} generates fewer traces than tracing \jit{} but more
than tracing \jit{} with a high threshold.

The second breakdown (the right-hand side of Figure~\ref{fig:rq1_num_ops_trace}: number of
operations in traces) shows
that two-level \jit{} generates fewer operations in traces than tracing \jit{} but
slightly more than tracing \jit{} with a high threshold. In addition, the ratio of
traces generated from threaded code overall number of traces are approximately 10\%.
These results stem from the characteristic of the lightweight compiler that it generates
compact traces during early execution phases.

These results demonstrate the ability of the two-level \jit{} to balance warm-up speed and trace
generation; lightweight compilation using threaded code can improve early performance.
Therefore, two-level \jit{} effectively works interpreter execution, threaded code
generation, and tracing together.

\subsection{Q2: How Good is the Peak Performance of Two-Level Compilation?}

\begin{figure}[!t]
  \centering
  \includegraphics[width=.9\columnwidth]{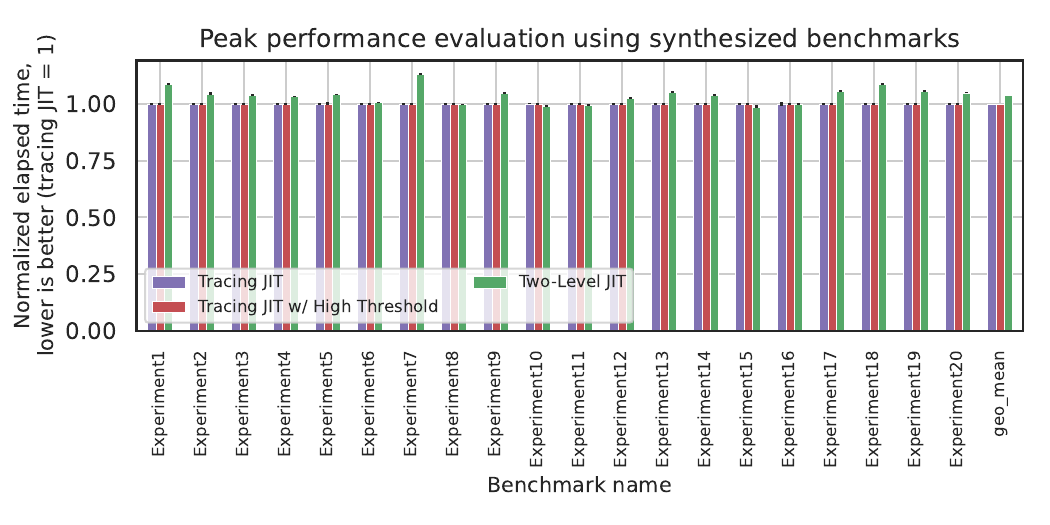}
  \caption{Peak performance results: Normalized elapsed time (lower is better) across
    synthesized
    programs for different JIT compilation strategies: (1) Tracing JIT, (2) Tracing JIT
    with a high threshold, and (3) Two-Level JIT. Results are normalized to the baseline
    Tracing JIT.}
  \label{fig:rq2}
\end{figure}

\begin{figure}[!t]
  \centering
  \includegraphics[width=.9\linewidth]{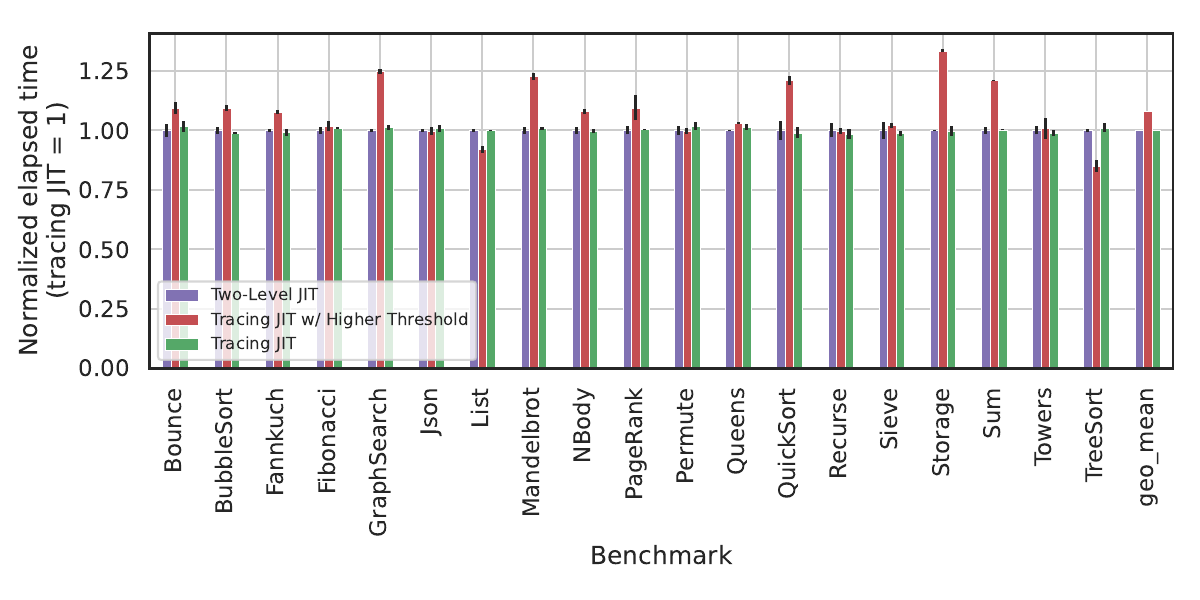}
  \caption{Peak performance results: Normalized elapsed time (lower is better) for SOM benchmarks,
  comparing three JIT compilation strategies: (1) Two-Level JIT, (2) Tracing JIT with a higher threshold, and
  (3) Tracing JIT. The results are normalized to the baseline Tracing JIT.}
  \label{fig:rq2_micro}
\end{figure}

In this section, we answer Q2 by describing the peak performance of two-level \jit{}
in 2SOM. We measure the peak performance of synthesized programs and the SOM
microbenchmark programs, and these results are shown in Figures~\ref{fig:rq2}
and~\ref{fig:rq2_micro}.

As shown in Figure~\ref{fig:rq2}, the peak performance of two-level \jit{} is slower
than approximately 3\% against tracing \jit{}, and approximately 5\% slower than
tracing \jit{} with high threshold. In addition, as shown in
Figure~\ref{fig:rq2_micro}, the peak performance of two-level \jit{} is slower than
approximately 7\% against tracing compilation and approximately 10\% slower than
tracing \jit{} with high threshold.

This result indicates that introducing the \tcg{} to a meta-tracing \jit{} compiler does not
significantly reduce the peak performance of meta-tracing \jit{} compilation.
In particular, the introduction of interpreter switching does not result in a
significant performance degradation in peak performance.

\subsection{Q3: How Much is the Improvement of Threaded Code Generation?}

\begin{figure}[t]
  \centering
  \includegraphics[width=\linewidth]{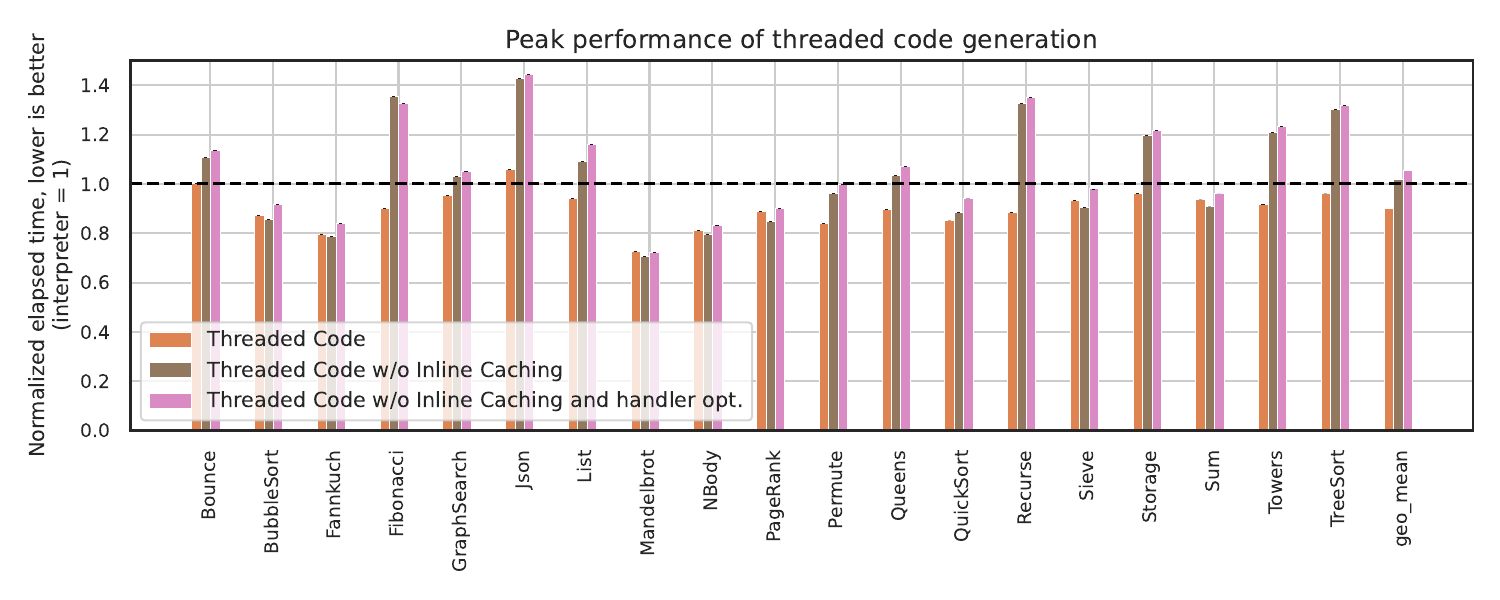}
  \caption{Comparison of peak performance for threaded code generation under different
    optimization configurations: (1) all optimizations enabled, (2) without inline
    caching, and (3) without both inline caching and shallow tracing handler
    optimization. Each data is normalized to interpreter-only execution.}
  \label{fig:rq3}
\end{figure}

In this section, we answer Q3 by describing the peak performance of the \tcg{}.
To evaluate the impact of the optimizations implemented in threaded code generation, we
measured its peak performance. Specifically, we compare three configurations: one with
all optimization options enabled, one with only the direct calls with inline caching
optimization disabled, and another with both direct calls with inline caching and shallow
tracing handler optimization disabled. This comparison allowed us to assess the
performance improvements contributed by each optimization individually. The result is
shown in Figure~\ref{fig:rq3}.

As shown in Figure~\ref{fig:rq3}, the fully optimized configuration (Threaded Code,
orange-colored bar) consistently achieves the best performance across all benchmarks,
running about 10\% faster than the interpreter. Disabling direct calls with inline caching
(brown-colored bar) degrades
performance by about 10\% on average, highlighting its importance in optimizing frequent
function calls and dynamic dispatch. Disabling both direct calls with inline caching and
shallow tracing handler optimization (pink-colored bar) further degrades performance by
about 5\%, highlighting their complementary impact on efficiency. Overall, the results
underscore the need for these optimizations to achieve optimal runtime performance.

This result shows that our implemented optimization improves the performance of the
\tcg{}. In particular, direct calls with inline caching significantly improve the
performance in function call-intensive programs such as Fibonacci, Json, Recurse, Storage,
Towers, and TreeSort. However, further optimization is needed to improve the warm-up
performance more. For example, we believe that inline caching implemented this time can be
optimized more efficiently by using polymorphic inline caching.

\subsection{Threats to Validity}

Since this evaluation was conducted on a subset of Smalltalk, the results
may differ when it is performed on a larger language implementation such as PyPy
and RSqueak~\cite{felgentreff:2016}.

\section{Related Work}
\label{sec:relatedwork}

There exist several approaches to building a lightweight compiler in a meta-compiler
framework. Copy-and-patch compilation~\cite{xu:2021:10.1145/3485513} is a meta-compilation
technique that performs template-based compilation. The template is generated from a bytecode
instruction or an AST node of a high-level language by the MetaVar compiler. Although they
build a new meta-compiler to generate a new baseline compiler, our approach is built on an
existing meta-tracing compiler. In addition, although they use a parameterized binary code
for generating a template, we use an interpreter definition with annotations to compile
a source program.

There are also other methodologies for realizing a multi-tier \jit{} compilation \vm{} using a meta-compiler
framework. \truffle{}~\cite{wurthinger:2017:ppe:3062341.3062381} has introduced
a first-tier execution in addition to the second-tier execution that performs heavyweight
compilation~\cite{prokopec:2021:graalmulti}. The approach is straightforward: they control the
level of optimization in their meta-compilation system to enable multi-tier execution.
While their methodology requires major modifications to the meta-compiler, our
methodology is adding multiple tiers with different compilation methods in \rpython{}.

Furthermore, other meta-circular \vm{}s provide multi-tier \jit{} compilation, such as
Jikes
RVM~\cite{arnold:2000:10.1145/353171.353175} and Maxine
\vm{}~\cite{wimmer:2013:10.1145/2400682.2400689}. These two \vm{}s have lightweight
and heavyweight compilers. In particular, they have original code generators for
lightweight compilation: Jikes RVM's lightweight compiler translates Java bytecode into
native code by simulating Java's operand stack. Maxine \vm{}'s lightweight compiler
translates Java functions into native code using templates. In contrast, our lightweight
compilation reuses the compiler of heavyweight compilation. To control the code quality
and compilation time, we slightly modify the definition of an interpreter provided to the
meta-tracing compiler.

\section{Conclusion and Future Work}
\label{sec:conclusion}

This paper introduces a lightweight methodology that enables a meta-compiler framework to generate
a multi-tier \jit{} compilation \vm{}. By treating interpreter definitions as specifications not only
for language semantics but also for compilation strategies, we enable \rpython{} to generate key components
for multi-tier \jit{} compilation: multiple compilers, a profiler, and a transition mechanism.

To demonstrate our approach, we created a two-tier \jit{} compilation version of the Simple Object Machine
called 2SOM. 2SOM incorporates two \jit{} compilers: a tier-1 threaded code generator and a tier-2 tracing
\jit{} compiler. This integration was achieved by addressing practical challenges in combining
meta-compiler-based threaded code generation~\cite{izawa:2022:threadedcodegen} with RPython.

Our evaluation shows that the proposed methodology significantly improves performance.
Warm-up performance was improved by approximately 15\%, while peak performance degradation was limited to
approximately 5\% compared to tracing \jit{}-only systems. Furthermore, enhancements to the threaded code generator
improved execution speed by 10\% compared to interpreter-based execution. These results demonstrated the
potential of our methodology to simplify the development of multi-tier JIT compilers in other RPython-based language implementations.

Despite its advantages, our approach has certain limitations.
First, it cannot add a tier that requires different low-level code generation and
optimization mechanisms (e.g., register allocation or instruction-level scheduling) because it
does not change the processes of \rpython{} after generated traces.
Second, it cannot add a tier that requires manipulating intermediate code. This
limitation arises because the approach relies on running the interpreter to obtain traces of
the intermediate code (RPython traces). The design does not assume modifications to these traces.

Future work is not only to solve these limitations, but also to apply our method to larger language
systems than 2SOM and validate the effectiveness of multi-tier JIT compilation VMs using more
realistic data.

One potential application is PyPy. Although we have conducted experiments using synthesized
programs to evaluate the effectiveness of multi-tier \jit{} compilation \vm{}s. However,
we think that using a larger benchmark set would enable more detailed analysis. To achieve this
in PyPy, two specific challenges must be addressed:
a) How to reduce the effort required to annotate PyPy’s interpreter, which consists of
approximately 200,000 lines of code.
b) How to handle cases where handlers in the PyPy interpreter may throw exceptions,
given that our current tier-1 compiler does not support tracing handlers that can raise RPython-level exceptions.
c) How to develop a lightweight compiler that can generate higher-quality code than
threaded code generation, enabling the multi-tier \jit{} compilation \vm{} to be effective for a
broader range of workloads.

For challenge a), it will be necessary to allow at least two interpreters to be generated from a
single interpreter. One idea is to improve PyPy’s translation flow, which generates VMs, to
automate the generation of interpreters.

For challenge b), we believe that adopting the \emph{zero-cost exception mechanism}~\cite{zerocostexceptionhandling}
introduced in Python 3.11--now being integrated into PyPy--will allow us to trace handlers in the PyPy interpreter.
Unlike earlier versions of Python where try statements compiled to \verb|SETUP_FINALLY| (incurring overhead even without exceptions),
Python 3.11 uses \verb|NOP| instructions for the non-exception path. This change should let our threaded code generation avoid
triggering RPython-level exceptions and work smoothly in PyPy with this mechanism enabled.

For challenge c), we believe that a solution can be achieved through further refinement of meta-compiler--based
threaded code generation. In general, interpreters implement the behavior of handler functions by
calling various helper functions within those handlers. The current threaded code generator only
leaves call instructions to the handler functions in the trace, but more efficient lightweight
compilation could be achieved by inlining non-critical helper functions and leaving only those
helper functions that are time-consuming to compile as call instructions.



\bibliography{main}

\ifextendedversion
\appendix
\section{Construction and Validation of Synthesized Benchmark Programs}
\label{app:construct_large_workload_app}

In this section, we describe the methodology to construct synthesized benchmark programs
used in Section~\ref{sec:evaluation} and discuss the characteristic of them through \jit{}
indicators through PyPy benchmark programs.

\subsection{Construction of a Large-Workload Program for Evaluating Two-Level Compilation}
\label{app:construct_based_on_dacapo}

The challenge in evaluating two-level compilation in 2SOM is how to
test our methodology on a large-scale application. However, the code base of
programs that 2SOM can run is smaller than that of Java or JavaScript \vm{}, and there is no
predefined benchmark set equivalent or similar to
DaCapo~\cite{blackburn:2006:10.1145/1167473.1167488} or
Renaissance~\cite{prokopec:2019:10.1145/3359061.3362778}. Therefore, we compose a
program that replicates a large application with non-trivial workloads.

Given this situation, our options are (1) implementing a large-scale benchmark set
comparable to DaCapo or Renaissance in the SOM language or (2) reproducing
the workload of a large-scale application using SOM's existing benchmark set.
While (1) would be the preferable approach, the SOM language lacks certain language
features, which are not used for SOM's existing benchmark set but are needed for implementing
a large-scale benchmark set, such as network communication and asynchronous processing,
necessitating extensions to the language itself. In addition, implementing a benchmark
set of this scale would require significant effort. For these reasons, we opt for approach (2).

To replicate the workload of a large-scale application, we investigated the correlation
between the number of method invocations and their rank.
The rank refers to the position in a descending order of method invocation
counts, where rank 1 indicates the most frequently invoked method, and the
lowest rank corresponds to the method that was called only once. We measured the
number of method invocations using the HPROF profiling
tool\footnote{\url{https://docs.oracle.com/javase/8/docs/technotes/samples/hprof.html}}
with interpreter execution for the DaCapo benchmark. Benchmarks with more than
400 method invocations were selected and plotted on logarithmic $x$- and
$y$-axes in Figure~\ref{fig:dacapo_corelation}.
The $y$-axis represents the number of method calls, while the $x$-axis
represents the rank of methods sorted in descending order based on their number
of calls. As shown in Figure~\ref{fig:dacapo_corelation}, regression analysis
revealed a high correlation  with an $R^2 > 0.98$. Therefore, we can conclude
that the program in this experiment should be structured such that the
correlation between method invocations and rank is high.

Based on the investigation, we design the experiment by combining all benchmark programs used
in Section~\ref{sec:evaluation}. The programs that comprise the experimental
setup are termed subprograms. These subprograms are executed continuously in a
single process. The number of
internal iterations assigned to the subprograms is adjusted so that the $R^2$ value for the
linear regression on the logarithmic scale of the number of method invocations and rank is as
close to 0.98 as possible. Figure~\ref{fig:exp_corelation} illustrates the correlation in the same manner as Figure~\ref{fig:dacapo_corelation}.

The experimental program consists of 20 subprograms. These subprograms are executed
sequentially from top to bottom, with each subprogram assigned a predetermined number of
iterations within the experimental program.
To generate the set of 20 experiment programs, the execution order of the subprograms is
randomly shuffled. Specifically, a base subprogram order is first determined. Then, by
shuffling the execution order while keeping the workload unchanged, an alternative
subprogram sequence is created. This process is repeated 19 more times to generate the
remaining 19 experimental programs.

\begin{figure}[t]
  \centering
  \begin{minipage}[t]{.475\linewidth}
    \centering
    \includegraphics[width=\linewidth]{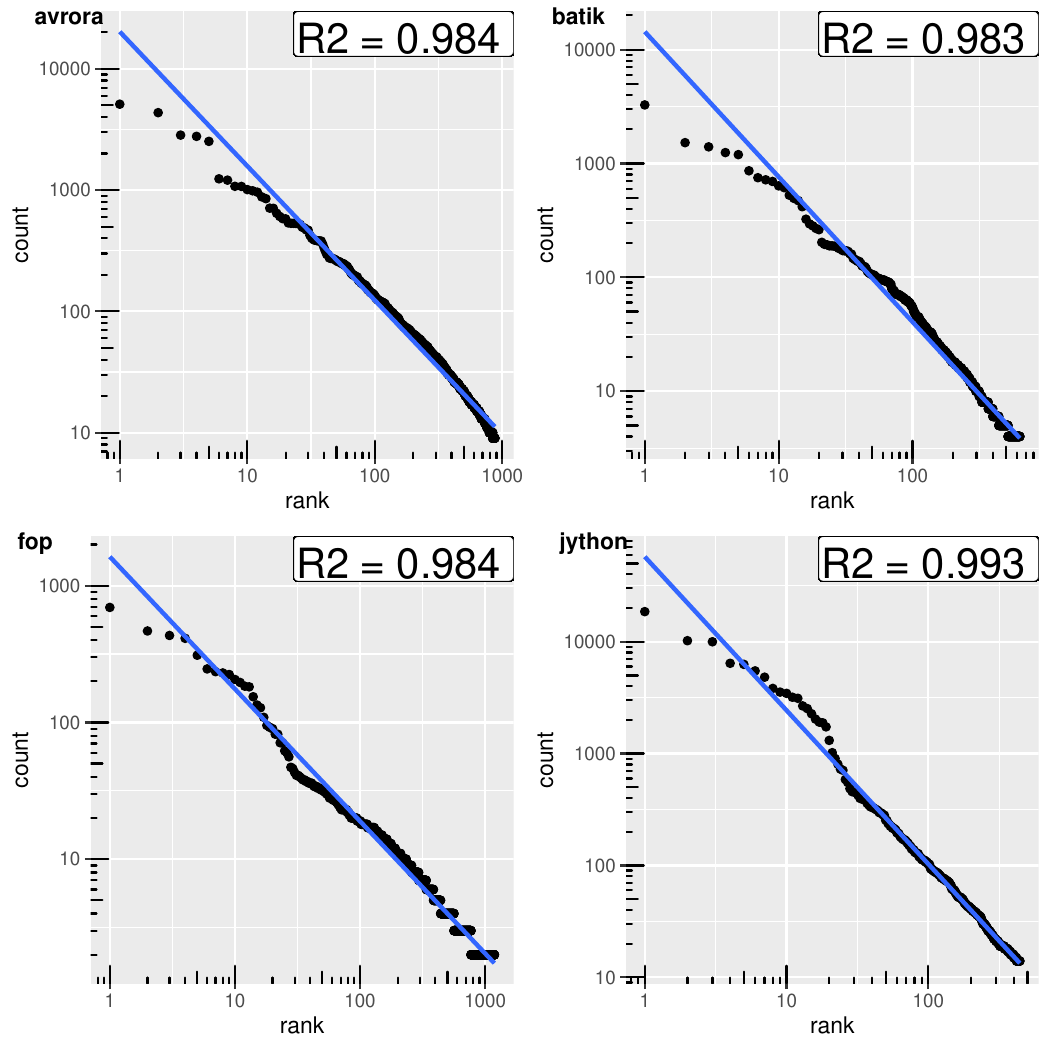}
    \caption{Correlation between the number of method invocations and rank for
        avrora, batik, fop, and jython in the DaCapo benchmark. $x$- and $y$-axes
        are logarithmic. The blue line is fitted by least-squares regression.}
    \label{fig:dacapo_corelation}
  \end{minipage}
  \hfill
  \begin{minipage}[t]{.475\linewidth}
    \centering
    \includegraphics[width=\linewidth]{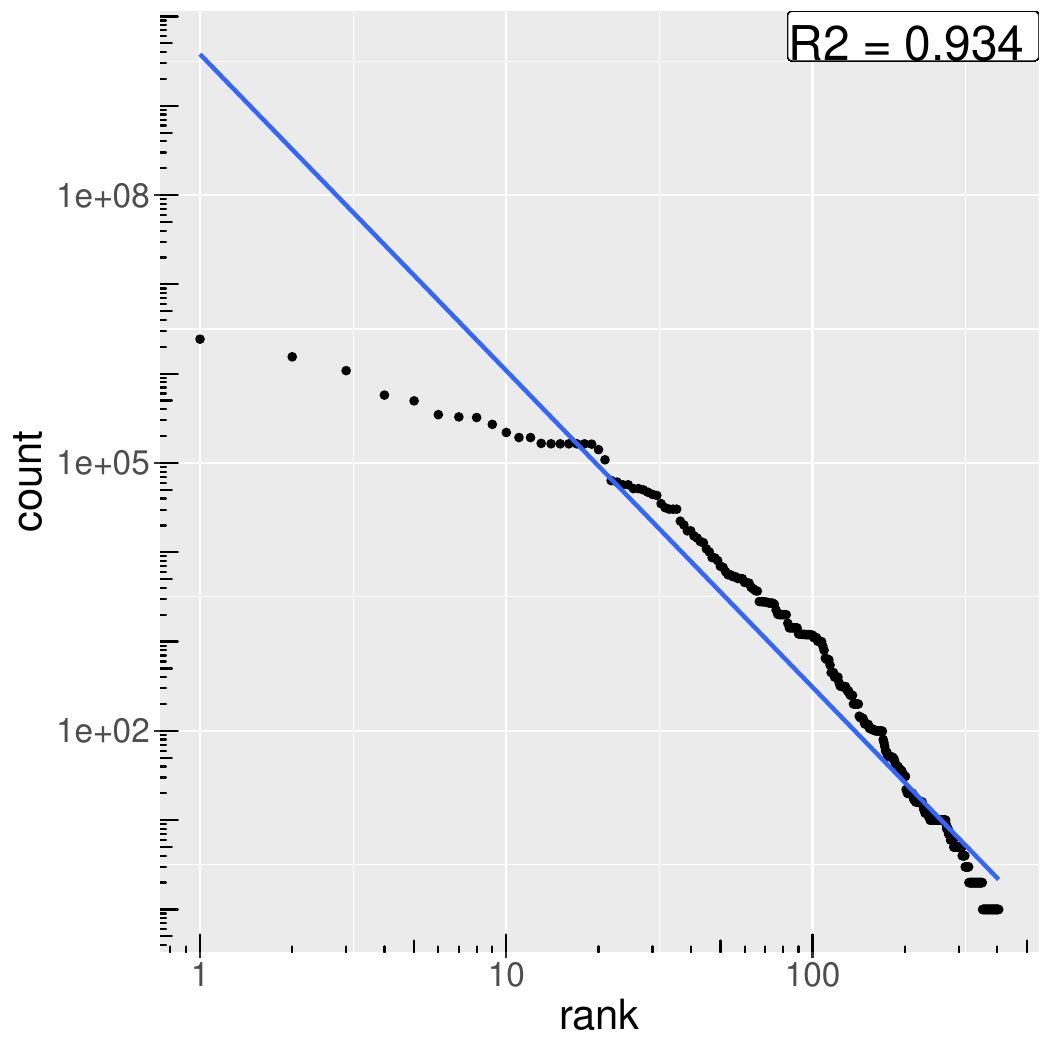}
    \caption{Correlation between the number of method invocations and rank for the program
      constructed for the experiment of multi-tier \jit{} compilation.}
    \label{fig:exp_corelation}
  \end{minipage}
\end{figure}

\subsection{Validation of Our Synthesized Benchmark with PyPy's Real-World Benchmark Program}
\label{app:comparison_pypy}

We validate the characteristics of our synthesized benchmark through PyPy's
benchmark programs, which are hosted at
Heptapod\footnote{\url{https://foss.heptapod.net/pypy/benchmarks}}.

First, we measure the correlation between the number of method invocations and
the rank of them as we measure the Dacapo benchmark.
We selected the top eight PyPy benchmarks with the largest number of executed
methods and conducted measurements on them.

The results are shown in Figure~\ref{fig:pypy_method_count_rank}. The $x$-axis
represents the method rank, and the $y$-axis represents the number
of method invocations; both axes use a logarithmic scale. The black dots
indicate the measured values, while the blue lines show the results of linear
regression. The $R^2$ value for each plot is displayed in the top right
corner. These results suggest that the benchmarks in the PyPy suite with a large
number of method invocations exhibit a distribution of method call frequencies
that closely resembles that of the Dacapo benchmarks.

\begin{figure}[t]
  \includegraphics[width=\linewidth]{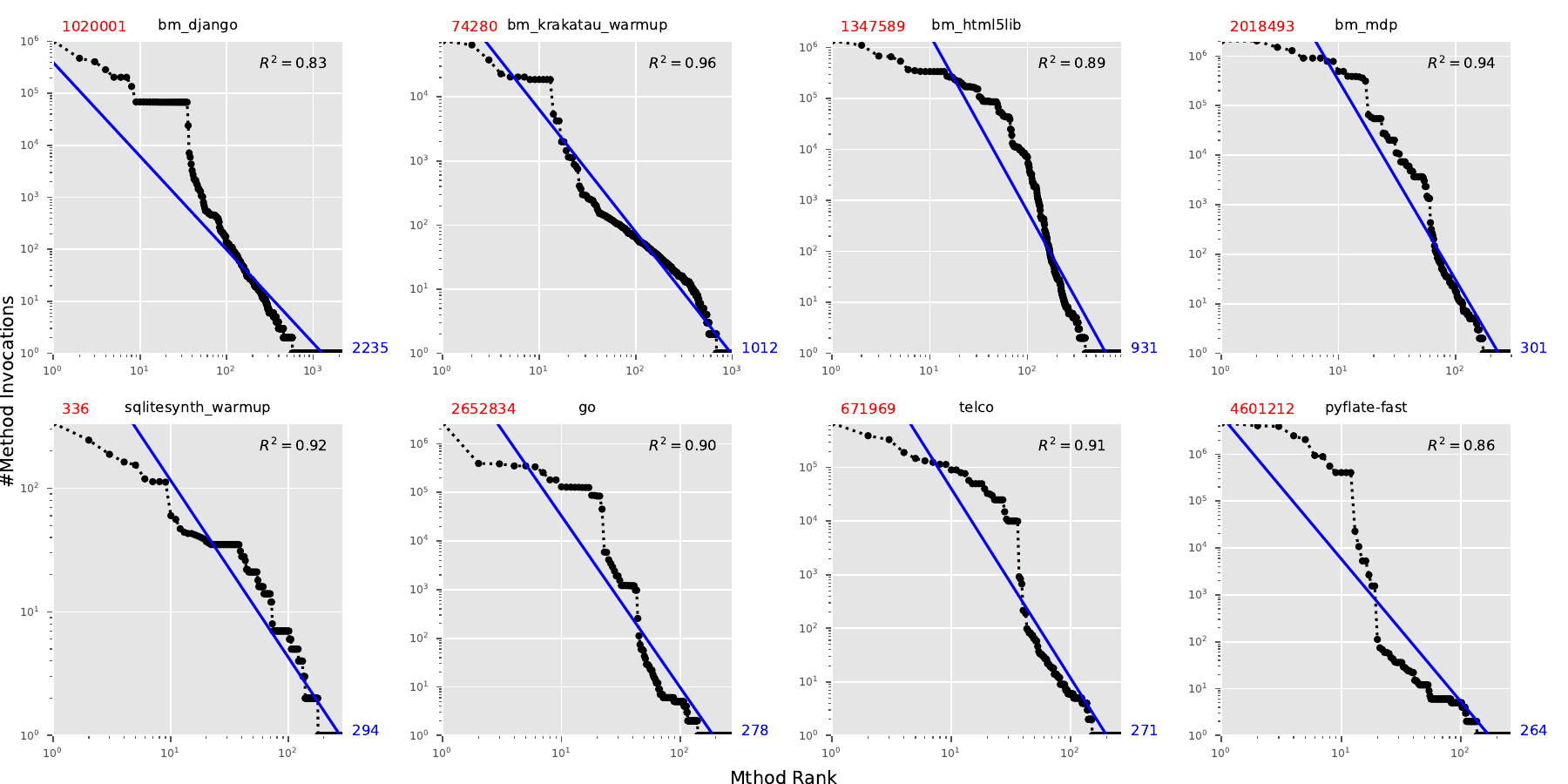}
  \caption{Correlation between the number of method invocations and the rank of
    methods.}
  \label{fig:pypy_method_count_rank}
\end{figure}

Next, we compare our synthesized benchmark with the PyPy benchmark programs and discuss
the characteristics of the synthesized workloads. Specifically, we use profiling
results from the tracing JIT compiler of PyPy/RPython for comparison. For the
PyPy benchmarks, we use PyPy's tracing JIT compiler, and for 2SOM, we consider
only the tracing JIT compiler. Using these setups, we run both the PyPy and
synthesized benchmarks to measure the average trace length (in terms of IR-level
instructions generated by the tracing JIT compiler) and examine its correlation
with the number of invoked methods.

While Figure~\ref{fig:pypy_method_count_rank} focused on the top eight
benchmarks with the most invoked methods, this analysis includes smaller
benchmarks as well, in order to capture broader trends and investigate the
characteristics of larger-scale benchmarks in the PyPy suite.

\begin{figure}[t]
  \centering
  \includegraphics[width=\linewidth]{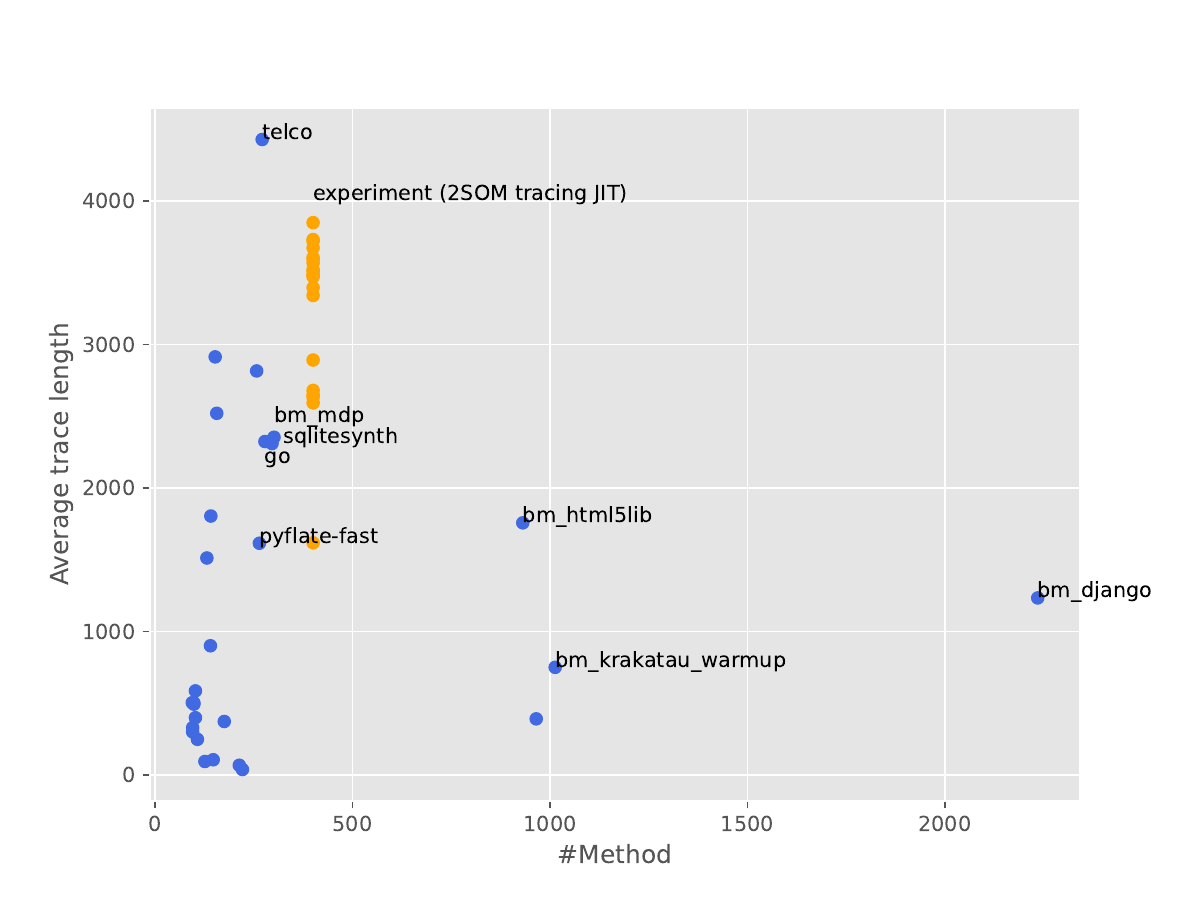}
  \caption{Correlation between average trace lengths and number of invoked methods
    in PyPy and synthesized benchmarks.}
  \label{fig:pypy_2som_ave_trace_len}
\end{figure}

The results are shown in Figure~\ref{fig:pypy_2som_ave_trace_len}. Blue dots
represent PyPy benchmarks, while orange dots represent the synthesized
benchmarks from 2SOM. For the top eight benchmarks by number of methods, their
names are annotated in the figure.

From the results, we observe that the synthesized benchmarks exhibit a
relatively constant number of hotspots, lacking the distribution observed in
PyPy benchmarks. Nevertheless, they do not deviate significantly from the
larger-scale PyPy benchmarks. For instance, \verb|telco|, \verb|bm_mdp|,
\verb|sqlitesynth|, and pyflate-fast are located near the synthesized
benchmarks. On the other hand, benchmarks with more than approximately 1,000
methods-—such as \verb|bm_html5lib|, \verb|bm_krakatau|, and
\verb|bm_django|-—are positioned far from the synthesized ones.

Thus, while the synthesized benchmarks may not capture the full diversity of
workloads, their overall scale is comparable to that of the larger PyPy
benchmarks.


\fi

\end{document}